\documentclass[fleqn,usenatbib]{mnras}

\usepackage[T1]{fontenc}
\usepackage[nomath]{stix2}
\usepackage{newtxmath}

\DeclareRobustCommand{\VAN}[3]{#2}
\let\VANthebibliography\thebibliography
\def\thebibliography{\DeclareRobustCommand{\VAN}[3]{##3}\VANthebibliography}

\usepackage{graphicx}
\graphicspath{{figures/}}
\usepackage{subcaption}
\usepackage{amsmath}
\usepackage{multirow}
\usepackage{tabularx}
\usepackage{balance}
\usepackage{soul}
\setstcolor{red} 
\setul{}{1.5pt}
\soulregister\citep7

\newcommand{\hinvMpc}{\,h^{-1}{\rm Mpc}}

\title[Colour- and luminosity-dependent CIC]{The conditional colour--magnitude distribution -- III. A study of galaxy colour- and luminosity-dependent counts-in-cells statistics}

\author[R. Isaac and Z. Zheng]{
Raymon Isaac\thanks{E-mail: \href{mailto:raymon.isaac@utah.edu}{raymon.isaac@utah.edu}}
and Zheng Zheng\thanks{E-mail: \href{mailto:zhengzheng@astro.utah.edu}{zhengzheng@astro.utah.edu}}
\\
Department of Physics and Astronomy, University of Utah, E2108 Stewart Building, 270 S 1400 E, Salt Lake City, UT 84112, USA
}

\date{Accepted XXX. Received YYY; in original form ZZZ}

\pubyear{\the\year{}}

\begin{document}
\label{firstpage}
\pagerange{\pageref{firstpage}--\pageref{lastpage}}
\maketitle



\begin{abstract}
\label{sec:abstract}
The conditional colour–magnitude distribution (CCMD) model connects the galaxy colour-luminosity distribution with dark matter halo mass, which has been constrained in Xu et al. by jointly modelling the two-point correlation functions of a large galaxy sample of the Sloan Digital Sky Survey (SDSS) in fine bins of galaxy colour and luminosity. We test the CCMD model of the galaxy–halo connection using counts-in-cells (CIC) statistics by comparing CCMD model predictions with SDSS measurements for 16 subsamples of galaxies in various luminosity and colour bins. In addition, we compare SDSS measurements with three analytic models of CIC distribution. The CCMD model provides a good description of the SDSS CIC measurements for most subsamples. The exceptions are the void probability functions and CIC distributions with cell radius $5\hinvMpc$ for a few subsamples in the luminous bins, which are not well described by the analytic models, either. For these subsamples, in general, removing the Sloan Great Wall from SDSS or introducing velocity bias to the CCMD model does not resolve the disagreement. Finally, with the help of the CCMD mocks, we investigate the cause of the variations of the galaxy counts in cells. For small cells or cells with a low mean galaxy count, the variation reflects effects from fluctuations in both the matter environment and the halo occupation distribution, while for large cells or cells with a high mean galaxy count, the environment effect dominates. We discuss the implications of our work for investigating galaxy assembly bias.

\end{abstract}




\begin{figure*}
    \centering
    \begin{subfigure}[b]{0.30\textwidth}
        \includegraphics[width=\linewidth]{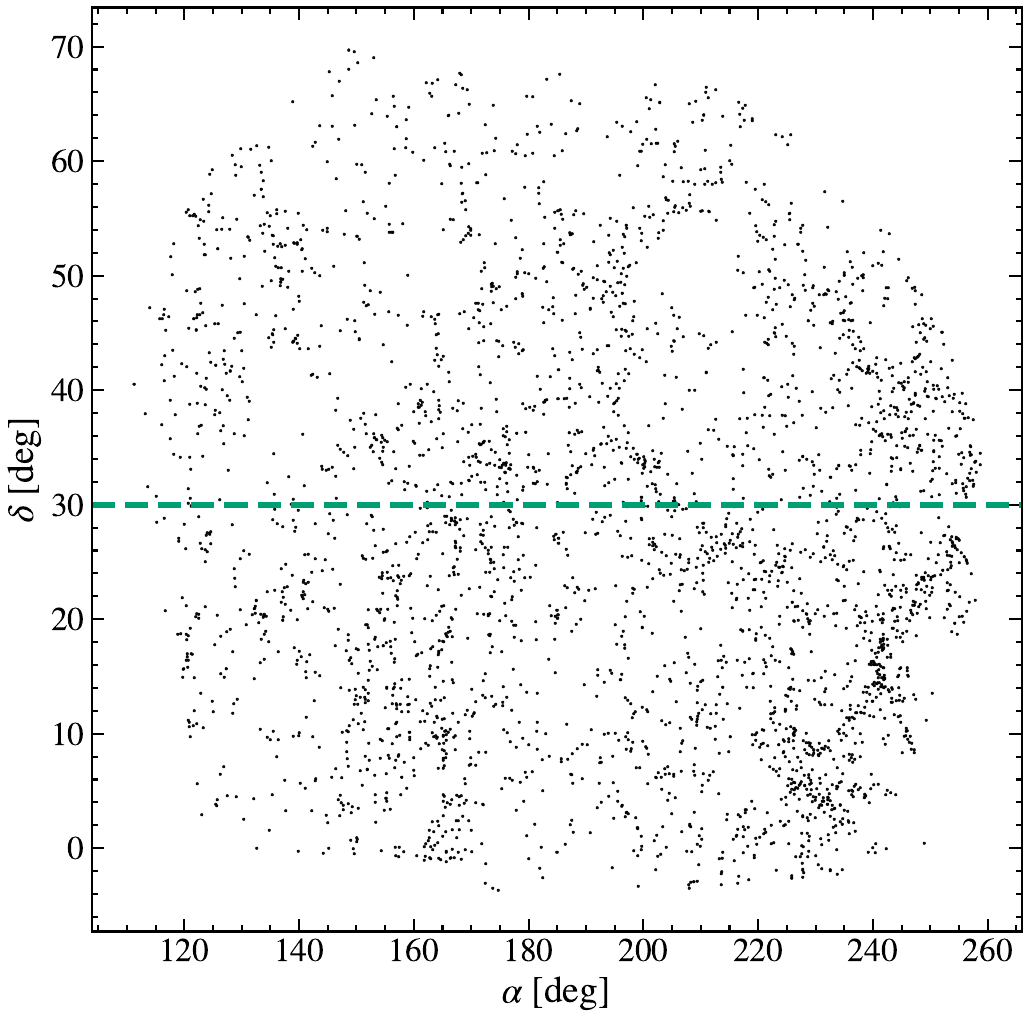}
        \label{fig:sdss_footprint_a}
    \end{subfigure}
    \begin{subfigure}[b]{0.30\textwidth}
        \includegraphics[width=\linewidth]{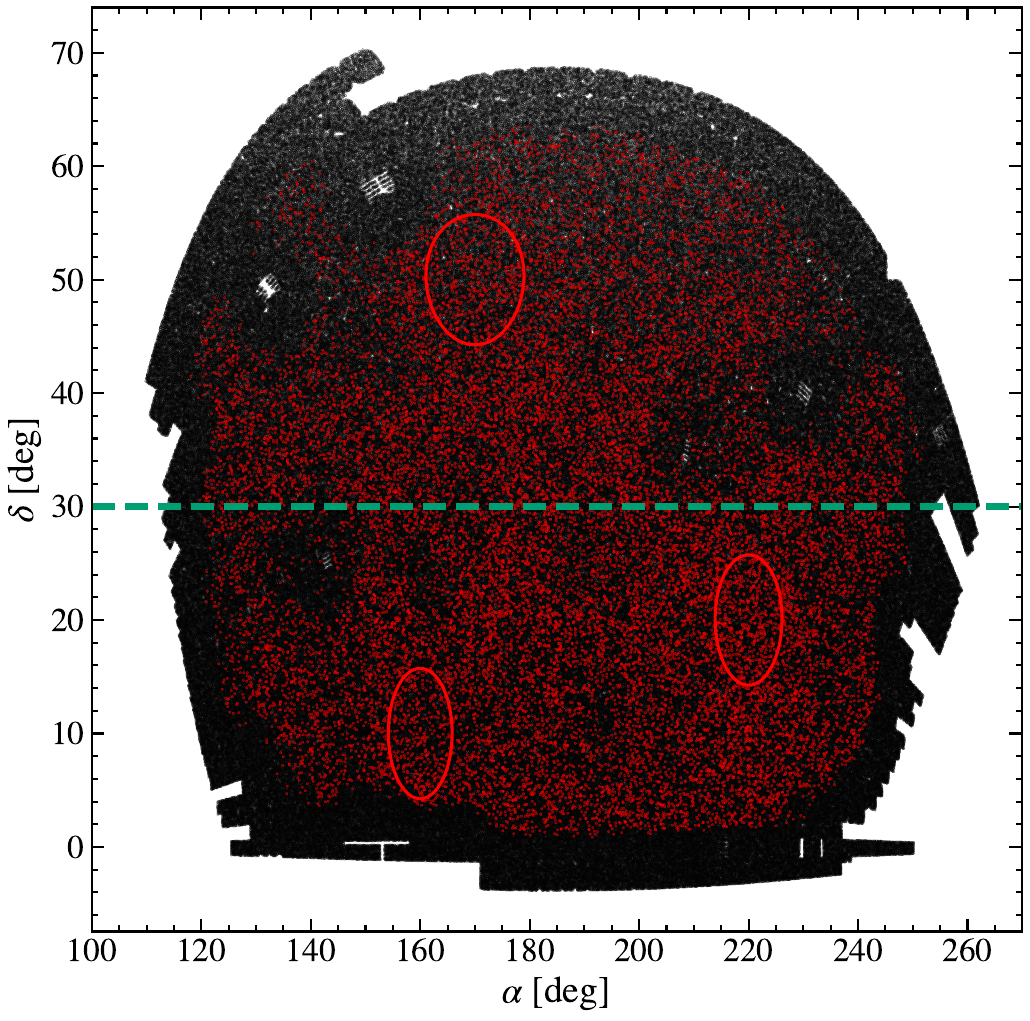}
        \label{fig:sdss_footprint_b}
    \end{subfigure}
    \raisebox{-5mm}{%
    \begin{subfigure}[b]{0.38\textwidth}
        \includegraphics[width=\linewidth]{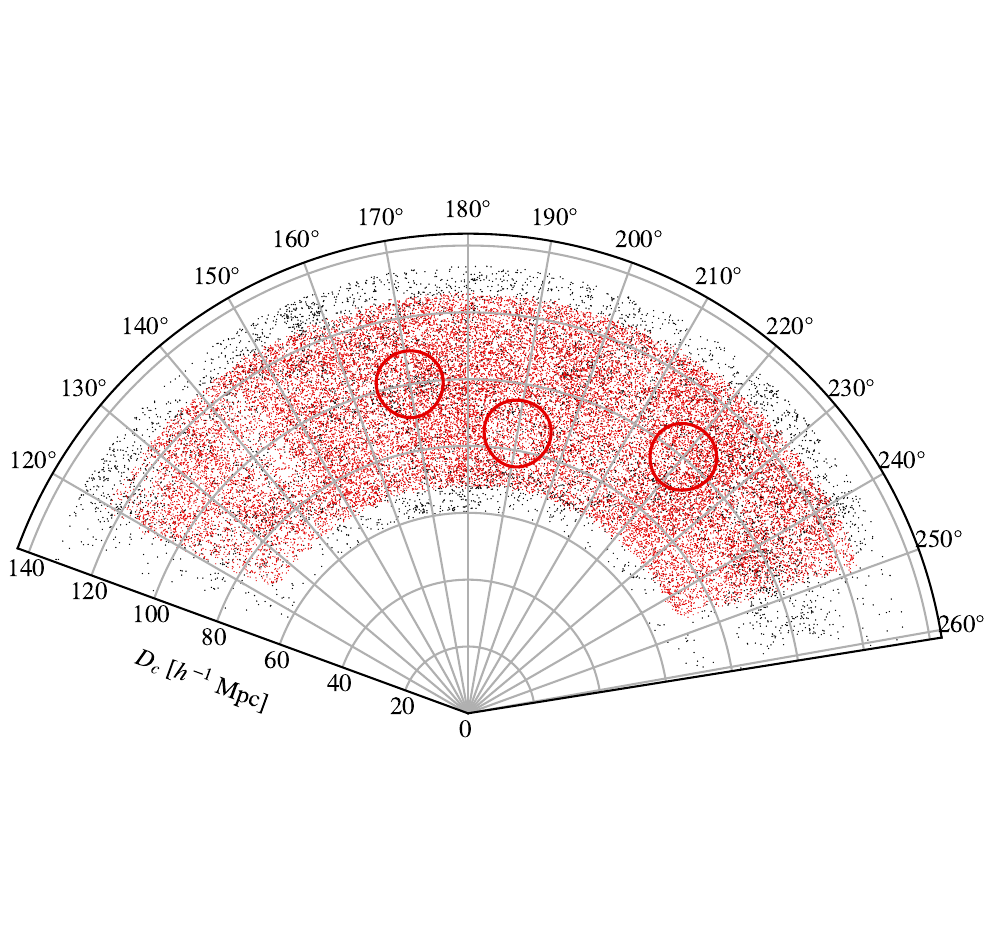}
        \label{fig:sdss_polar}
    \end{subfigure}
    }
    \caption{An illustration of distributions of galaxies and randomly placed cells for the SDSS subsample with $-19 \leq M_\mathrm{r} \leq -18$ and $0.00 \leq g-r \leq 0.50$. The left panel shows the sky distribution of galaxies in the redshift range of $0.03<z<0.04$. In the middle panel, the black points are drawn from a random catalogue to show the SDSS footprint and the masked regions. The red points are the centres of randomly placed cells of radius $R = 10 \hinvMpc$, with the three circles being examples of the cells. These centres are chosen to be within a slice of comoving distance of $D_{\rm c}=100\pm 1\, \hinvMpc$. The placements of cells account for the edge of the footprint and masked regions. The right panel is a polar projection around $\delta=30^\circ$ (as indicated by the dashed lines in the left two panels) of the SDSS subsample (black points) and the centres of the randomly placed cells (red points), with three examples of the cells (red circles).}
    \label{fig:sdss_footprint}
\end{figure*}

\section{Introduction}
\label{sec:Introduction}

In the standard model of galaxy formation and evolution \citep[e.g.][]{White1978MNRAS.183..341W}, galaxies form and evolve inside dark matter haloes. One of the goals of modern astrophysics is to characterise the relationship between observable properties of galaxies and the properties of their host haloes. This galaxy–halo connection (see \citealt{Wechsler2018ARA&A..56..435W} for a review) can reveal the physics of galaxy formation, and it can also help cosmological constraints from large galaxy surveys, where galaxies are used as tracers of the underlying dark matter distribution.

Luminosity and colour are the most fundamental and informative qualities of galaxies \citep{Blanton2005ApJ...629..143B} that are directly observable. In the colour–magnitude diagram (CMD), galaxies show a bimodal distribution, which appears in the local galaxy population as well as high-redshift galaxy populations \citep[e.g.][]{Strateva2001AJ....122.1861S, Baldry2004ApJ...600..681B, Faber2007ApJ...665..265F}. It contains a blue cloud of actively star-forming, disc-dominated galaxies and a red sequence of quiescent, spheroidal galaxies, separated by a sparsely populated green valley. The existence and evolution of this bimodality are driven by many physical processes that quench star formation, including internal ones, such as feedback-driven outflows, and external ones, such as ram-pressure stripping or strangling in dense environments \citep{Gunn1972ApJ...176....1G, Peng2015Natur.521..192P}. Studying the galaxy-halo connection in terms of galaxy luminosity and colour would help understand the origin and evolution of the bimodal distribution of galaxies.

Empirical approaches within the halo model \citep[e.g.][]{Cooray2002} have been effective in inferring the galaxy-halo connection based on observed galaxy clustering. The halo occupation distribution (HOD; e.g. \citealt{Jing1998ApJ...494....1J, Seljak2000MNRAS.318..203S, Berlind2002ApJ...575..587B, Zheng2005ApJ...633..791Z}) and the conditional luminosity function (CLF; e.g. \citealt{Yang2003MNRAS.339.1057Y}) are two widely adopted frameworks. Applications of these frameworks have successfully constrained the tight relationship between galaxy luminosity and halo mass from modelling galaxy clustering statistics [e.g. the two-point correlation function (2PCF)] in large galaxy surveys \citep[e.g.][]{Zehavi2005ApJ...630....1Z, Zehavi2011ApJ...736...59Z, vandenBosch2013MNRAS.430..725V}. Various parameterizations and methods have been introduced to model the colour dependence of galaxy clustering, in addition to the luminosity dependence, to constrain the relation between galaxy colour and halo mass \citep[e.g.][]{Zehavi2005ApJ...630....1Z, Skibba2009MNRAS.392.1080S, Zehavi2011ApJ...736...59Z}.

A comprehensive model capable of describing the joint distribution of galaxy colour and luminosity conditioned on halo mass, dubbed the conditional colour–magnitude distribution (CCMD), is proposed by \cite{Xu2018MNRAS.481.5470X}. It represents a substantial and sophisticated extension of the CLF framework, incorporating the colour dimension to provide a complete probabilistic description of the galaxy population within the three-dimensional space of colour, luminosity (absolute magnitude), and halo mass. The CCMD model has been constrained through jointly modelling the colour- and luminosity-dependent 2PCFs of a large sample of galaxies in fine bins of colour and luminosity from the Sloan Digital Sky Survey (SDSS; \citealt{York2000AJ....120.1579Y, Strauss2002AJ....124.1810S, Blanton2005AJ....129.2562B, Abazajian2009ApJS..182..543A}).

In this paper, we aim to test the CCMD model by analysing the luminosity- and colour-dependent counts-in-cells (CIC) statistics of galaxies. The CIC statistics describe the probability distribution of the number of galaxies of given properties in cells of various sizes, which effectively probe the higher-order spatial clustering of galaxies \citep[e.g.][]{White1979MNRAS.186..145W,Saslaw2000dggc.conf.....S, Saslaw2010LectureNotes}. In particular, it is shown in \citet{White1979MNRAS.186..145W} that the void probability function (VPF) is related to the complete hierarchy of connected $N$-point correlation functions and provides the generating function for the full distribution of the clustering, making it sensitive to galaxy clustering beyond the 2PCF. As a powerful probe of the galaxy distribution, the CIC statistics can also help constrain the galaxy--halo connection beyond the dependence on halo mass \citep[e.g.][]{Wang2019}, i.e. galaxy assembly bias. We will measure the CIC statistics using SDSS data and compare them with predictions from the CCMD model constrained by the 2PCFs \citep{Xu2018MNRAS.481.5470X}. In addition, we will compare the SDSS measurements with three widely used analytic CIC distributions \citep[e.g.][]{Yang2011ApJ...729..123Y,Hurtado-Gil2017A&A...601A..40H}.

The paper is organized as follows. In Section~\ref{sec:Data}, we describe the SDSS data used for the CIC measurements and the CCMD mocks constructed based on 2PCF modelling. In Section~\ref{section:counts-in-cells measurements and observational models}, we describe the methodology and procedure to measure CIC statistics and introduce the analytic models used in this study. In Section~\ref{section:Results}, we present the results of CIC analyses. Finally, we summarize the results in Section~\ref{section:Summary and Discussion}.


\section{Data}
\label{sec:Data}

In this study, we compare the luminosity- and colour-dependent CIC measurements from the SDSS data with those from the galaxy mock catalogue based on the CCMD model. We also use the CCMD mocks to help us understand the CIC statistics in terms of the distribution of haloes and halo occupation of galaxies. In this section, we describe the SDSS data and the CCMD mock catalogues, respectively.


\subsection{SDSS DR7 data}
\label{subsection:SDSS DR7 Data}

We use the galaxy data from the SDSS Data Release 7 (DR7; \citealt{Abazajian2009ApJS..182..543A}). In particular, our galaxy subsamples are constructed based on the large-scale structure sample {\tt dr72bright1} of the New York University Value-Added Galaxy Catalogue (NYU-VAGC; \citealt{Blanton2005AJ....129.2562B})\footnote{\url{https://sdss.physics.nyu.edu/vagc/}}. The sample has a legacy spectroscopic footprint area of 8032 deg$^2$. In this paper, we exclude the discrete stripes from the continuous main area because, in many cases, they are narrower than the cell size we use. The reduced footprint area becomes $\sim 7139$ deg$^2$ (see Fig.~\ref{fig:sdss_footprint}). The properties of galaxies used in this study include $r$-band absolute magnitude $M_{\rm r}$, $g-r$ colour, right ascension $\alpha$, declination $\delta$, and redshift $z$.

\begin{figure}
    \centering
    \hspace*{-0.48cm}
    \begin{subfigure}[b]{0.422\textwidth}
        \includegraphics[width=\linewidth]{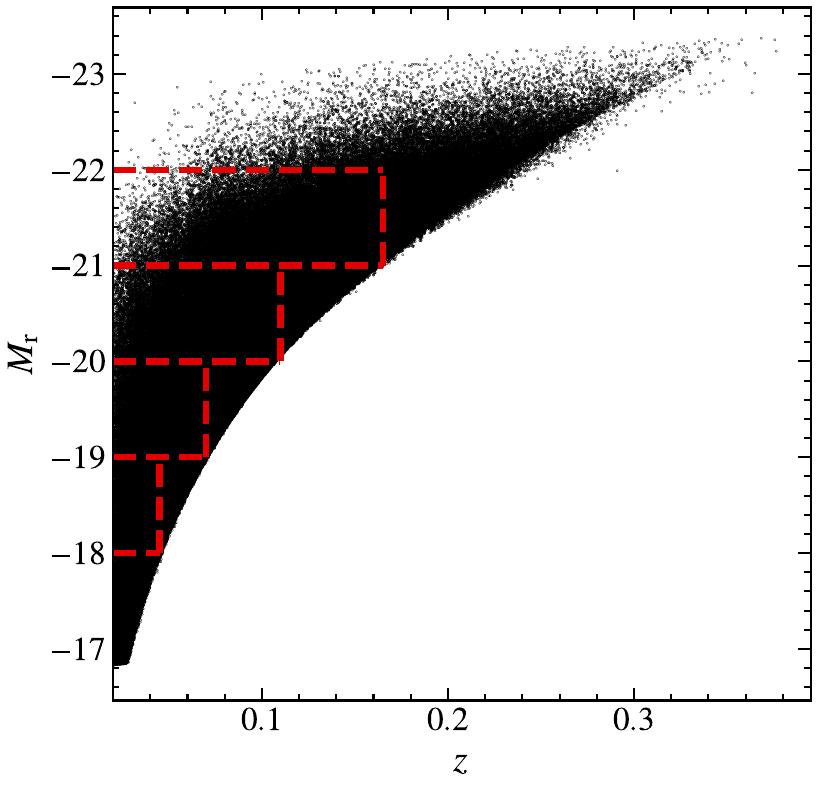}
    \end{subfigure}
    \begin{subfigure}[b]{0.43\textwidth}
        \includegraphics[width=\linewidth]{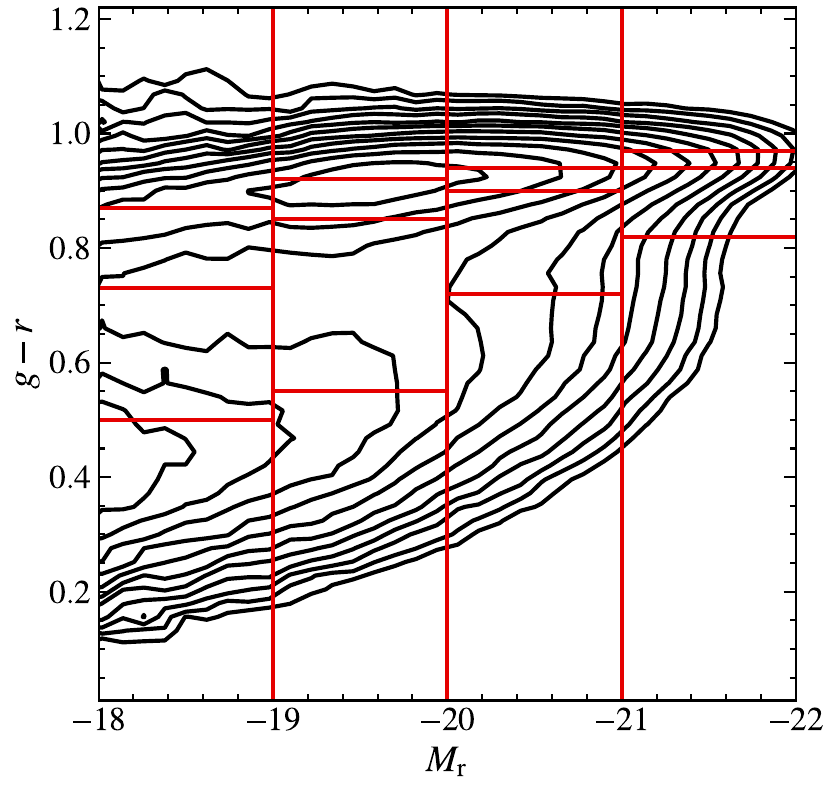}
    \end{subfigure}
    \caption{Luminosity and colour subsamples. In the top panel, the black points show the distribution of the absolute magnitude and redshift of galaxies. Volume-limited subsamples in four luminosity bins are illustrated by the red boxes, all with a minimum redshift $z_{\rm min}=0.02$. In the bottom panel, the galaxy CMD is shown. The three vertical lines separate galaxies into four luminosity bins, and in each luminosity bin, the three horizontal lines separate galaxies into four colour subsamples. In total, we have 16 luminosity and colour subsamples of galaxies.}
    \label{fig:redshift_cut}
\end{figure}

Based on the large-scale structure sample, we construct luminosity and colour subsamples. We first divide the main sample into four volume-limited luminosity samples depending on the magnitude $M_\mathrm{r}$, with $\Delta M_\mathrm{r} = 1$ in the magnitude range of $-22 \leq M_\mathrm{r} \leq -18$. We set the minimum redshift $z_\mathrm{min} = 0.02$ for all samples. For each luminosity sample, the maximum redshift $z_\mathrm{max}$ is chosen to make the sample volume-limited. The top panel of Fig.~\ref{fig:redshift_cut} shows the redshift cuts of the four luminosity samples. Then, for each luminosity bin, we further divide the sample into four subsamples depending on the galaxy colour, which has a range of $0.00 \leq g-r \leq 1.20$. The bottom panel of Fig.~\ref{fig:redshift_cut} shows the colour cuts of the 16 subsamples in the CMD, overlaid with the CMD of galaxies. In each magnitude bin, the four colour subsamples have a comparable number density of galaxies. We end up with 16 galaxy subsamples across the CMD, and their information is listed in Table~\ref{tab:sample_information}.

We also make use of the 20 random catalogues associated with the large-scale structure sample for CIC measurements as described in \ref{subsection:Counts-in-Cells Measurements}. Each random catalogue includes the right ascension $\alpha$ and the declination $\delta$ of $\sim 2\times 10^6$ random points. For subsamples in each magnitude bin, we randomly assign the radial comoving distances up to $z_{\rm max}$ to the random points so that they have a uniform number density within the corresponding volume. 

We note that all the magnitudes and colours are $K-$ and evolution-corrected to the median redshift of galaxies in the SDSS DR7 main sample, $z \sim 0.1$. The magnitude is calculated by setting $h = 1$, where $h$ is the Hubble constant in units of 100 ${\rm km\, s^{-1} Mpc^{-1}}$. In this paper, the terms ``luminosity'', ``magnitude'', and ``absolute magnitude'' are used interchangeably. 

\begin{table}
    \centering
    \caption{Luminosity and colour subsamples of galaxies. For each subsample, listed are the magnitude range, the colour range, the maximum redshift, the total number of galaxies, and the mean number density of galaxies. The minimum redshift is set to be $z_{\rm min}=0.02$ for all subsamples.
    }
    \label{tab:sample_information}
    \begin{tabularx}{8.5cm}{cccrc}
        \hline
        $M_\mathrm{r}$ (mag) & $g-r$ (mag) & $z_\mathrm{max}$ & \multicolumn{1}{c}{$N_\mathrm{g}$} & \shortstack{$n_\mathrm{g}$ \\ $\left[10^{-3} h^{3} \mathrm{Mpc}^{-3}\right]$}\rule{0pt}{20pt}\\
        \hline
        \multirow[t]{4}{*}{$[-18, -19]$} & $[0.00, 0.50]$ & \multirow[t]{4}{*}{0.045} & 7294 & 4.636\rule{0pt}{10pt}\\
                                       & $[0.50, 0.73]$ &                           & 5466 & 3.474\\
                                       & $[0.73, 0.87]$ &                           & 3309 & 2.104\\
                                       & $[0.87, 1.20]$ &                           & 3148 & 2.002\\
        \multirow[t]{4}{*}{$[-19, -20]$} & $[0.00, 0.55]$ & \multirow[t]{4}{*}{0.070} & 13794 & 2.212\\
                                       & $[0.55, 0.85]$ &                           & 20537 & 3.294\\
                                       & $[0.85, 0.92]$ &                           & 9601 & 1.540\\
                                       & $[0.82, 1.20]$ &                           & 9762 & 1.566\\
        \multirow[t]{4}{*}{$[-20, -21]$} & $[0.00, 0.72]$ & \multirow[t]{4}{*}{0.110} & 35646 & 1.490\\
                                       & $[0.72, 0.90]$ &                           & 32901 & 1.375\\
                                       & $[0.90, 0.94]$ &                           & 17602 & 0.736\\
                                       & $[0.94, 1.20]$ &                           & 23907 & 0.999\\
        \multirow[t]{4}{*}{$[-21, -22]$} & $[0.00, 0.82]$ & \multirow[t]{4}{*}{0.165} & 29413 & 0.378\\
                                       & $[0.82, 0.94]$ &                           & 28375 & 0.366\\
                                       & $[0.94, 0.97]$ &                           & 13921 & 0.179\\
                                       & $[0.97, 1.20]$ &                           & 14941 & 0.192\\
        \hline
    \end{tabularx}
\end{table}


\subsection{CCMD mock catalogues}
\label{subsection:CCMD Mock Catalogues}

The CCMD model \citep{Xu2018MNRAS.481.5470X} is a global model that describes the distribution of galaxy luminosity and colour as a function of halo mass. Motivated by the bimodal distribution of galaxy colour, the model introduces two populations of galaxies, pseudo-red and pseudo-blue galaxies, with each population separated into central and satellite galaxies. 

The CCMD for central galaxies of each pseudo-colour population is represented by a two-dimensional Gaussian distribution, and that for satellite galaxies is characterised by a modified Schechter function for luminosity distribution and a Gaussian function for colour distribution. The parameters for the central and satellite CCMD distributions are all functions of halo mass. The CCMD model is constrained by simultaneously fitting the number densities and projected 2PCFs of a large sample ($\sim 80$) of SDSS DR7 galaxies in fine bins of colour and luminosity \citep{Xu2018MNRAS.481.5470X}.

The CCMD modelling results have been tested beyond the 2PCFs. In \citet{Xu2024MNRAS.533.1485X}, by applying the same galaxy group finding algorithms to SDSS galaxies and to those in a CCMD mock catalogue, galaxy colour and luminosity distributions in galaxy groups are compared between model predictions and data, as a function of group mass. The comparisons demonstrate a remarkable consistency between the CCMD groups and the SDSS groups. In this work, we extend the comparison and test to the CIC statistics through CIC measurements in the CCMD mock catalogue and in the SDSS data.

The overall CCMD mock catalogue\footnote{\url{https://zhengzheng-astro.github.io/data.html}} is constructed by populating galaxies into haloes identified in the MultiDark MDPL2 $N$-body cosmological simulation \citep{Klypin2016MNRAS.457.4340K}, which adopts a spatially flat Planck cosmology \citep{Planck2014A&A...571A..16P, Planck2016A&A...594A..13P} with $\Omega_\text{m}$ = 0.307 and $\Omega_\text{b}$ = 0.048 and has a periodic box of 1000$\hinvMpc$ on a side. This is the same simulation used in \citet{Xu2018MNRAS.481.5470X} to constrain the CCMD parameters.
The best-fitting CCMD parameters are adopted in constructing the mock catalogue. In each halo, the central galaxy is assigned the position and velocity of the potential minimum of the halo, and satellite galaxies are assigned the positions and velocities of randomly chosen dark matter particles inside the halo. The CCMD galaxy mock catalogue includes galaxy ID,  position, velocity, $r$-band absolute magnitude $M_\mathrm{r}$, $g-r$ colour, pseudo-colour type, and central/satellite status. The associated halo catalogue includes halo ID, position, velocity, mass ($\log[M_\mathrm{h}/(h^{-1} \mathrm{M}_\odot)]$), number of galaxies hosted, and starting ID of the hosted galaxies. With the overall mock catalogue, we apply the same cuts in luminosity and colour as in the SDSS data to create mock catalogues for the 16 subsamples of galaxies.

The positions of galaxies and haloes in the CCMD galaxy and halo catalogues are in real space. For comparison with SDSS data, we choose the $Z$-direction as the line-of-sight direction to produce the redshift-space catalogues. Under the distant observer approximation,  the $Z-$component in redshift space is calculated as (for $z=0$) 
\begin{equation}
Z_\text{redshift} = Z + \frac{v_Z}{H_0},
\end{equation}
where $v_Z$ is the peculiar velocity in the $Z$-direction. For the purpose of investigating the covariance matrices of CIC statistics, we also produce cone-shaped catalogues centred around chosen positions, and we perform similar calculations for the position shift in redshift space along the line of sight of each galaxy.

       
\section{CIC measurements and analytic models}
\label{section:counts-in-cells measurements and observational models}

In this section, we present the CIC measurement method to be applied to the SDSS data and the CCMD mocks. In addition to comparing the SDSS measurements with the CCMD model, we also examine the agreement between the SDSS measurements and three commonly used analytic models of CIC statistics, which are described here.

\subsection{CIC measurements}
\label{subsection:Counts-in-Cells Measurements}

For a given subsample of galaxies, the CIC measurement depends not only on the clustering of galaxies but also on the number density of the subsample. To focus on comparing the clustering of galaxies between SDSS data and models, we choose to match the number densities of the SDSS data and the CCMD mock. For each subsample in the two low-luminosity bins, we take the lower number density between the CCMD mock and the SDSS data. For each subsample in the two high-luminosity bins, we take the lowest number density among the CCMD mock, the SDSS data, and the SDSS data excluding the Sloan Great Wall (SGW; \citealt{Gott2005ApJ...624..463G}). The SDSS data with SGW excluded will be used to perform tests, as discussed later. We perform random subsampling to reduce the number density of each catalogue to match the desired one, and the typical change in number density is less than 10 per cent.

With each SDSS galaxy subsample, we make CIC measurements by randomly placing spherical cells of fixed radius inside the sample volume. Cell radii are limited to $1.0 \leq R \leq 20.0$ $h^{-1}$Mpc, and we focus on presenting results on the VPF and CIC distribution with cell radius $R=5\hinvMpc$ and $20\hinvMpc$, respectively. 

We note that the CIC distribution (including the VPF) can be obtained from the cumulative distribution function (CDF) of $k$ nearest-neighbour galaxies around randomly chosen points \citep[e.g.][]{White1979MNRAS.186..145W,Banerjee2021MNRAS.500.5479B}. Such CDFs can be efficiently computed by constructing a $k$-d tree on the data \citep[][]{Banerjee2021MNRAS.500.5479B}. Although our method of directly counting galaxies in random cells is less efficient, given the moderate survey volume and number of galaxies in each subsample, it is still tractable with acceptable computational cost.

To ensure that randomly placed cells are within the sample volume and to account for survey masks, we make use of the 20 random catalogues associated with each SDSS subsample. For a given subsample, with each random catalogue, we count the number of random points within each of our randomly placed cells and exclude those cells with counts below the mean count. The process is repeated with the other random catalogues. As an illustration, the red points in the middle panel of Fig.~\ref{fig:sdss_footprint} show the centres of the selected cells on the sky for the $-19\leq M_{\rm r}\leq -18$ and $0.00\leq g-r\leq 0.50$ subsample, while those in the right panel show the cell centres on a thin slice along the radial direction around $\delta=30^\circ$ (delineated by the green dashed line in the left and middle panels). Clearly, the selected cells avoid the boundaries of the footprint as well as masked regions within the footprint.

To determine the number of cells used in the measurements, following \citet{Conroy2006ApJ...647..201C}, we perform convergence tests for the mean number $\bar{N}$ of galaxies within the cells, the VPF, and the volume-averaged 2PCF $\bar{\xi}(V)$. The last quantity is calculated as 
\begin{equation}
\bar{\xi}(V) = \frac{\langle (\Delta N)^2 \rangle - \bar{N}}{\bar{N}^2},
\end{equation}
where $\Delta N=N-\bar{N}$ (number of galaxies within the cell minus the mean count). It results from the fractional difference between the mean count of galaxy pairs within cells and that from a Poisson distribution of mean $\bar{N}$. Based on the tests, we adopt a typical number of cells $\mathcal{N} \sim 10^5$ to measure $f_V(N)$, the probability of finding $N$ galaxies in a cell (with $N=0$ being the VPF).

To estimate the uncertainty in the CIC measurements, we apply the jackknife resampling method using $N_\mathrm{JK}=200$ jackknife subsamples. The jackknife subsamples are constructed based on the clustering of galaxies in right ascension $\alpha$ and declination $\delta$. The K-means method is employed to find 200 cluster centres, and galaxies are matched to the nearest cluster centre. As a consequence, the footprint is divided into 200 subregions of similar area.  The jackknife covariance matrix is calculated as
\begin{equation}
\text{Cov}(f_i, f_j) = \frac{N_\mathrm{JK}-1}{N_\mathrm{JK}} \sum_{l=1}^{N_\mathrm{JK}} \left( f_i^l - \bar{f}_i \right) \left( f_j^l - \bar{f}_j \right),
\end{equation}
where $\bar{f}_i$ is the mean value of the statistic $f_i$ in the $i$th bin and $f_i^l$ is the statistic measured with the $l$th subregion excluded. Similar to how we place random cells within the sample survey volume, we use the random catalogues to ensure that cells that are within or overlap with the $l$th subregion are excluded.
 
With the CCMD mock catalogue of each subsample (in redshift space), we perform the CIC measurements by randomly placing spherical cells with the periodic boundary condition accounted for. There are two types of uncertainty estimates. First, by dividing the box into 200 subregions, we apply the jackknife method to estimate the uncertainty. Such an estimate is on the box scale, a much larger scale than those corresponding to the SDSS sample volumes. It can be regarded as reflecting the CCMD model uncertainty. Second, for each subsample, we produce cone-shaped catalogues with the same angular area and maximum comoving radius as the corresponding SDSS subsample. With the limited size of the box, the number of independent cone-shaped catalogues we can construct depends on the volume of the corresponding SDSS subsample, with 81, 24, 3, and 3 cones for the subsamples of the lowest to the highest luminosity ($M_\mathrm{r}$) bin. For each subsample, we perform CIC measurements with each cone-shaped catalogue, and the covariance of the measurements coming from different cone-shaped catalogues forms the cone-based covariance matrix.

\subsection{Analytic CIC descriptions}
\label{subsection: Analytical Models}

In addition to the comparison between the CIC measurements from the SDSS data and the CCMD mocks, we compare the SDSS measurements with three analytic models. In previous work \citep[e.g.][]{Yang2011ApJ...729..123Y,Hurtado-Gil2017A&A...601A..40H}, these commonly adopted models have been found to provide reasonable descriptions of CIC measurements. Note that the exact form of the CIC distribution is determined by the full hierarchy of connected $N$-point correlation functions, as derived in previous work (e.g. \citealt{White1979MNRAS.186..145W,Szapudi1993ApJ...408...43S,Banerjee2021MNRAS.500.5479B}; also see \citealt{Banerjee2022MNRAS.511.2765B} for the exact form of the first nearest neighbour CDF, equivalent to the VPF). The analytic models adopted here only include the 2PCF and neglect higher-order clustering. While they can break down in certain cases, these simple models remain useful to fit CIC measurements.

\subsubsection{The gravitational quasi-equilibrium distribution}
\label{subsubsection:The Gravitational Quasi-Equilibrium Distribution}

On sufficiently large scales, the distribution of galaxies is nearly uniform, with the CIC distribution approaching the Poisson distribution. However, on smaller scales, gravity-induced clustering leads to deviations from the Poisson distribution. As a modification to the Poisson distribution, the gravitational quasi-equilibrium distribution (GQED) is introduced by \cite{Saslaw1984ApJ...276...13S} based on fluid thermodynamics of galaxies. It can also be derived from statistical mechanics, as shown in \cite{Ahmad2002ApJ...571..576A}. 

The GQED function has the form
\begin{equation}
f_{V}(N) = 
\frac{\bar{N}(1-k)}{N!} \,
[\bar{N}(1-k) + Nk]^{N-1} \,
\exp[-\bar{N}(1-k) - Nk],
\end{equation}
where $N$ is the number of galaxies found in the cells, $\bar{N}$ is the mean number of galaxies per cell, and $k$ is the clustering parameter  \citep{Yang2011ApJ...729..123Y},
\begin{equation}
k = 1 - [\bar{N}\,\bar{\xi}(V) + 1]^{-\tfrac{1}{2}}.
\end{equation}
As $k \to 0$ (i.e. $\bar{\xi} \to 0$), the GQED approaches the Poisson distribution.

\subsubsection{The negative binomial distribution}
\label{subsubsection:The Negative Binomial Distribution}

The negative binomial distribution (NBD) is first used in a cosmological context by \cite{Carruthers1983PhLB..131..116C}.  \cite{Elizalde1992MNRAS.254..247E} derive the probability distribution $f_{V}(N)$ by considering the number of galaxies in disconnected boxes of constant volume $V$. It has the form
\begin{equation}
f_{V}(N) = 
\frac{\Gamma\!\left(N + \frac{1}{\bar{\xi}}\right)}{\Gamma\!\left(\frac{1}{\bar{\xi}}\right) \, N!} \,
\frac{\bar{N}^N \left(\frac{1}{\bar{\xi}}\right)^{\frac{1}{\bar{\xi}}}}{\left(\bar{N} + \frac{1}{\bar{\xi}}\right)^{N + \frac{1}{\bar{\xi}}}},
\end{equation}
where $\Gamma$ is the gamma function. As $\bar{\xi}=\bar{\xi}(V) \to 0$, the NBD converges to the Poisson distribution. While it is found that the NBD violates the second law of thermodynamics \citep[e.g.][]{Saslaw1996ApJ...460...16S, Yang2011ApJ...729..123Y}, it remains a commonly used model in describing CIC statistics.

\subsubsection{The lognormal distribution with bias}
\label{subsubsection:The Log-Normal Distribution with Bias}

While the initial cosmological density field is well described by a Gaussian random field, the field evolved under gravitational instability can be approximated by the lognormal distribution (LND), as first noticed by \citet{Hubble1934ApJ....79....8H} and formally proposed by \citet{Coles1991MNRAS.248....1C} for cosmological applications.

We adopt the distribution function in \citet{Arnalte-Mur2016JCAP...03..005A} with a modification by \citet{Hurtado-Gil2017A&A...601A..40H} to account for the bias $b$ in the galaxy density fluctuation $\delta_{\rm g}$ with respect to the matter density fluctuation $\delta_{\rm m}$. With $\delta_m$ following the LND and a linear bias relation $\delta_g=b\delta_m$, the LND + Bias model for the distribution of $\delta_{\rm g}$ has the form
\begin{equation}
f(\delta_\mathrm{g}) = 
\frac{1}{\sqrt{2 \pi \mathcal{H}_0} (\delta_\mathrm{g} + b)}
\exp \left(-\frac{1}{2}\frac{y^2}{\mathcal{H}_0}\right),
\end{equation}
with
\begin{equation}
y = \ln \left[\sqrt{1 + \frac{\sigma_\mathrm{g}^2}{b^2}}\left(1+\frac{\delta_\mathrm{g}}{b}\right)\right]
\end{equation}
and
\begin{equation}
\mathcal{H}_0 = \ln\left(1 + \sigma_\mathrm{m}^2\right),
\end{equation}
where $\sigma_\mathrm{m}^2$ and $\sigma_\mathrm{g}^2$ are the variances in matter and galaxy density fluctuations, respectively. For a spherical cell of radius $R$, $\sigma_{\mathrm{g}}^2$ is measured as
\begin{equation}
\sigma_{\mathrm{g}}^2
= \left\langle \left( \frac{N - \bar{N}}{\bar{N}} \right)^2 \right\rangle
- \frac{1}{\bar{N}},
\end{equation}
where the term $1/\bar{N}$ represents the variance from the Poisson process. For $\sigma_{\mathrm{m}}^2$, we calculate it theoretically as
\begin{equation}
\sigma_{\mathrm{m}}^2(R,z)
= \frac{1}{2\pi^2}
\int_0^{\infty} k^2 P(k,z) \left|\tilde{W}(kR)\right|^2 \mathrm{d}k,
\end{equation}
where $k$ is the wavenumber in comoving $h$Mpc$^{-1}$, $P(k,z)$ is the linear matter power spectrum at redshift $z$, and $\tilde{W}(kR)$ is the Fourier transform of a spherical top-hat filter of radius $R$,
\begin{equation}
\tilde{W}(kR)
= \frac{3}{(kR)^3}\left[\sin(kR) - kR \cos(kR)\right].
\end{equation}
Our calculation uses the {\tt Colossus} toolkit \citep{Diemer2018ApJS..239...35D} with the \cite{Eisenstein1998ApJ...496..605E} transfer function for $P(k,z)$. With a linear bias relation between and $\delta_m$, we obtain $b = \sigma_g/\sigma_m$. 

Since $\delta_g\ge -1$, we further normalize the distribution $f(\delta_g)$ through
\begin{equation}
\tilde{f}(\delta_g) = \frac{f(\delta_g)}{\int_{-1}^{\infty} f(\delta_g)\mathrm{d}\delta_g}.
\end{equation}
For randomly placed cells, they traverse all possible density fluctuations. The probability of finding $N$ galaxies in a cell is obtained by integrating over the distribution of galaxy density fluctuations,
\begin{equation}
f_V(N) = \int_{-1}^{\infty} 
P(N|\delta_g) \tilde{f}(\delta_g) \, \mathrm{d}\delta_g,
\end{equation}
where $P(N|\delta_g)$ is the probability of finding $N$ galaxies in a Poisson process given the galaxy density fluctuation $\delta_g$,
\begin{equation}
P(N|\delta_g) = 
\frac{\bigl[\bar{N}(1 + \delta_g)\bigr]^N 
\exp[-\bar{N}(1+\delta_g)]}{N!}.
\end{equation}


\section{Results}
\label{section:Results}

In this section, we first show the results for the VPF and the general CIC distribution. We make comparisons between the measurements from the SDSS data and the CCMD mocks. The performance of the three analytic CIC distributions in describing the SDSS measurements is also examined. Finally, with the help of the CCMD mocks, we investigate the relation between the CIC statistics and the underlying halo abundance and halo occupation function of galaxies.


\subsection{VPF}
\label{subsection:Void Probability Distribution}

\begin{figure*}
    \centering
    \includegraphics[width=\textwidth]{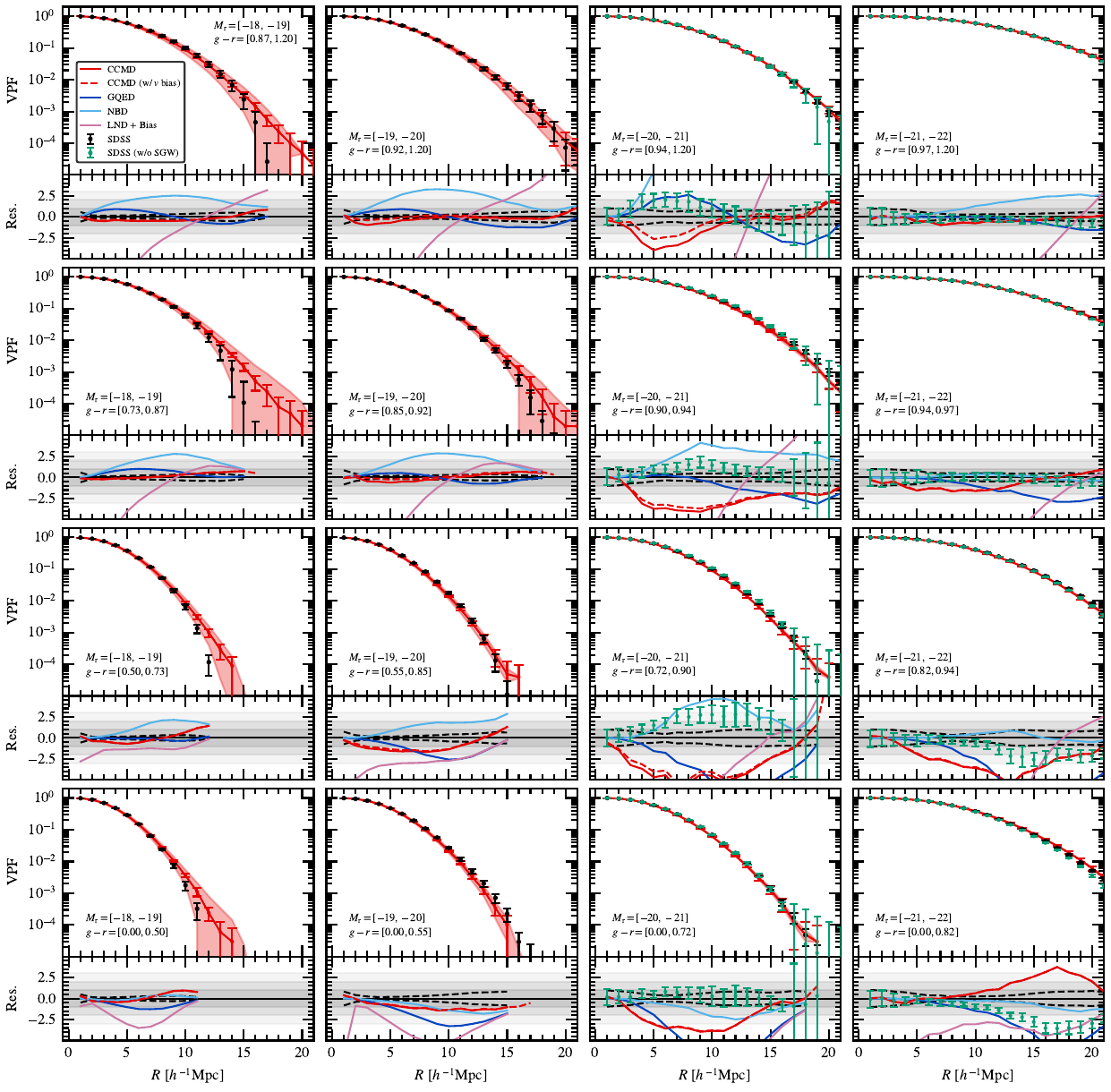}
    \caption{VPF measurements and model comparisons for the 16 subsamples of galaxies. The panels are arranged so that from left to right, galaxy luminosity increases and in each column, from bottom to top, galaxy colour changes from blue to red. In each main panel, the black points are the VPF measurements from the SDSS data with jackknife error bars. The red points are the VPF measured from the CCMD mock with jackknife error bars estimated within the whole box, while the red shaded region shows the uncertainty estimated from cone-shaped mocks that mimic the SDSS volume. For subsamples in the two luminous bins, VPF measurements from the SDSS data with the SGW excluded are shown as green points with jackknife error bars. In each residual panel, the dashed lines are the SDSS jackknife errors in terms of the total estimated errors (see the text for details). The curve for each model is the difference between the model VPF and the SDSS measurements, in units of the total estimated error. The shaded regions represent the 1$\sigma$, 2$\sigma$, and $3\sigma$ ranges. For subsamples in the two luminous bins, the green points in the residual panels are the differences between the VPF measurements without and with the SGW, in units of the total estimated error.
    }
    \label{fig:all_VPF}
\end{figure*}

We show the VPF $f_V(N=0)$ as a function of cell radius $R$, for the 16 luminosity and colour subsamples in Fig.~\ref{fig:all_VPF}. In each main panel, the black points with error bars represent the measurements from the SDSS data, while the red curves are from the CCMD mock of the corresponding subsample. 

The VPFs from the data and from the mocks all show the typical behaviour of being close to unity as the cell radius approaches zero and becoming exponentially small at large cell radii. There is an overall trend that the VPF has a higher amplitude toward redder subsamples at a fixed luminosity or toward more luminous subsamples at a similar colour. As the VPF depends on galaxy number density, we perform tests of VPF comparisons with number density matched. For example, with two subsamples,  we randomly sample the one with a higher number density to reduce its number density to match that of the other subsample and then compare the corresponding VPFs. With the number density matched, the above trend persists and becomes clearer. This trend is a manifestation that more luminous and redder galaxies are more strongly clustered, as seen in their 2PCFs \citep{Xu2018MNRAS.481.5470X}.

To quantify the agreement between the VPFs from the SDSS data and from the CCMD mocks, we first make an assessment of the error bars. The error bars for the SDSS measurements are estimated by using the jackknife resampling method. For VPFs from the CCMD mocks, two types of uncertainty are shown in each main panel of Fig.~\ref{fig:all_VPF}. The error bars on the red curve are obtained by applying the jackknife method to the cubic box of each mock, reflecting the sample variance on scales of the box size with the best-fitting CCMD model, and we treat them effectively as model uncertainty. The shaded regions are uncertainties estimated from the covariance among measurements with cone-shaped mocks that mimic the SDSS survey volume for each subsample. For most of the low-luminosity subsamples ($-20\leq M_{\rm r} \leq -18$), with respect to the uncertainties estimated from cone-shaped mocks, the uncertainties can be neglected, as the box of the mocks is much larger than the survey volumes of the corresponding SDSS subsamples. The uncertainties associated with the SDSS measurements are typically lower than those from the cone-shaped mocks, indicating that the SDSS measurement uncertainties from the jackknife resampling are underestimated. For high-luminosity subsamples,  the survey volumes become much larger, and each subsample can only have three independent cone-shaped mocks constructed, which makes the covariance matrix estimated from cone-shaped mocks noisier. The model uncertainty and that from the limited number of cone-shaped mocks become comparable. The uncertainties associated with the SDSS measurements are generally slightly higher.

Based on the assessment, for each subsample, we combine the covariance matrix from the SDSS measurements and that from the cone-shaped mocks to account for the underestimate of the jackknife method with the SDSS data for low-luminosity subsamples and the limitation of the small number of cone-shaped mocks for high-luminosity subsamples. Specifically, for each subsample, the diagonal elements of the two covariance matrices are added. With this sum, the correlation matrix from the SDSS covariance matrix is used to produce the total, representative covariance matrix that we use in the following analyses.

In each of the ``residual'' panels of Fig.~\ref{fig:all_VPF}, the difference between the VPF measurements with the CCMD mocks and with the SDSS data, in terms of the diagonal errors, is shown as the red curve. The shaded regions denote the levels of 1--3$\sigma$, and the dashed lines represent the fraction of the SDSS jackknife error in the total error budget. Comparisons between the SDSS measurements and the three analytic models (GQED, NBD, and LND + Bias) are also shown in the ``residual'' panels as blue, cyan, and purple curves. 

\begin{figure}
    \centering
    \includegraphics[width=0.45\textwidth]{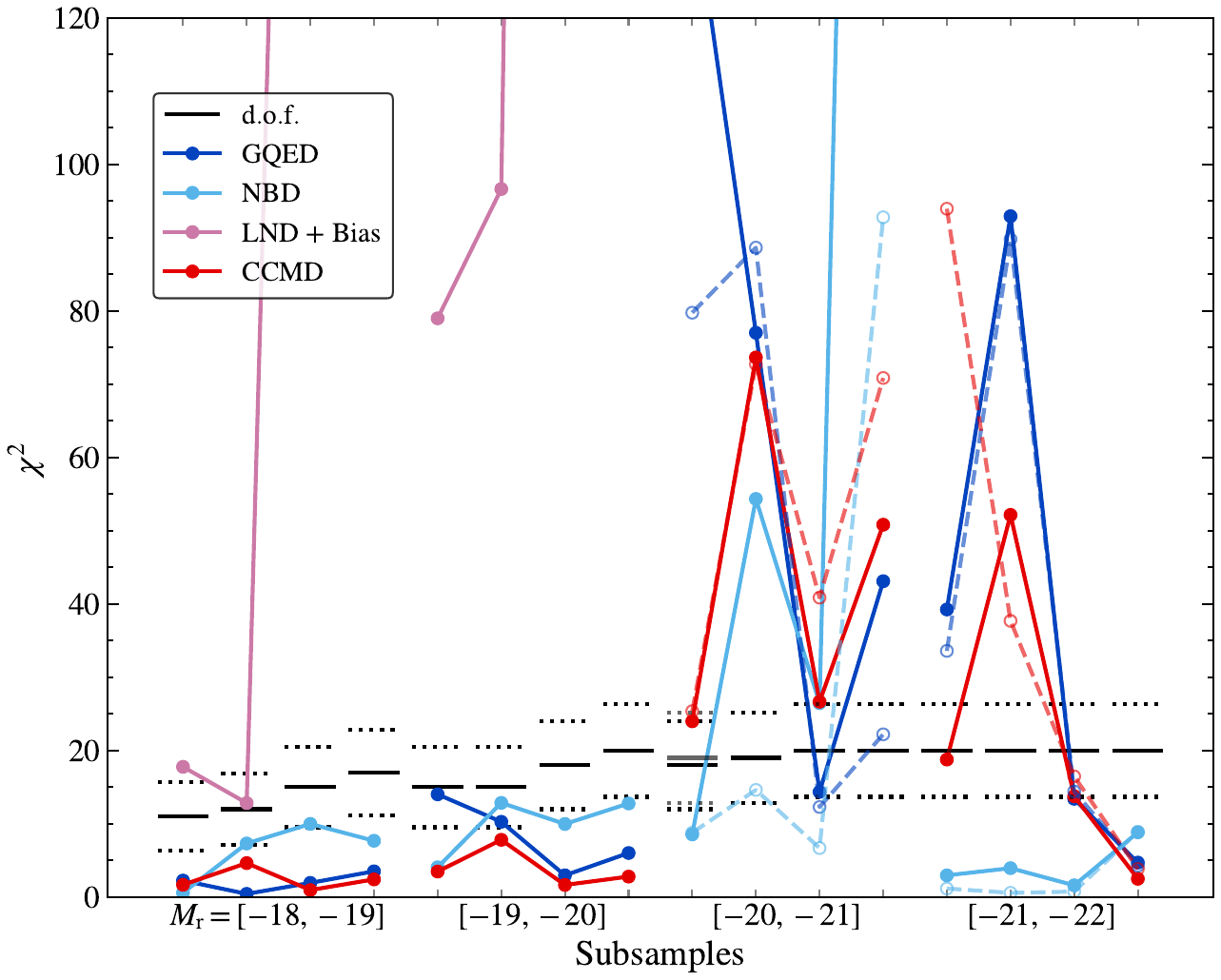}
    \caption{Values of $\chi^2$ for different models to describe the SDSS VPF measurements. The 16 subsamples are grouped into four luminosity bins, as marked along the horizontal axis. In each group, from left to right, the colour of the galaxies changes from blue to red. The solid (dashed) lines of different colours are $\chi^2$ values from different models in describing the SDSS VPF measurements with (without) the SGW. The solid and dotted black horizontal lines associated with each subsample illustrate the expected $\chi^2$ value and scatter, based on the d.o.f. of the subsample. The set of grey lines corresponds to those from the subsample with the SGW excluded. 
    }
    \label{fig:VPF_chi2_values}
\end{figure}

For either the CCMD model or any of the analytic models, we quantify the agreement with the SDSS measurements by calculating the $\chi^2$ value,
\begin{equation}
\chi^2 = (\mathbf{F}_{\rm SDSS}-\mathbf{F}_{\rm model})^\mathrm{T} \mathbf{C}^{-1} (\mathbf{F}_{\rm SDSS}-\mathbf{F}_{\rm model}),
\end{equation}
where $\mathbf{F}$ is the vector of the statistic (which is the VPF as a function of $R$ here), and $\mathbf{C}$ is the total covariance matrix. Table~\ref{tab:VPF_chi2} lists the degrees of freedom (d.o.f.) and the $\chi^2$ value calculated for each model for each subsample. The $\chi^2$ values for different models and different subsamples are also plotted in Fig.~\ref{fig:VPF_chi2_values} to show the general trend, with horizontal lines denoting the expected 1$\sigma$ range of the $\chi^2$ distribution, with mean = d.o.f. and scatter $\sim \sqrt{2{\rm d.o.f.}}$.

\begin{table*}
    \centering
    \caption{Values of $\chi^2$ of different models in describing the VPF measurements of 16 subsamples. For the two luminous bins, the subsamples include the SGW. The last column lists the d.o.f. for each subsample. 
    }
    \label{tab:VPF_chi2}
    \begin{tabularx}{12cm}{ccrrrrrc}
        \hline
        $M_\mathrm{r}$ (mag) & $g-r$ (mag) & GQED & \hspace{2mm}NBD & LND + Bias & CCMD & \shortstack{CCMD\\(w/ $v$ bias)} & d.o.f.\rule{0pt}{17pt}\\
        \hline
        \multirow[t]{4}{*}{$[-18, -19]$} & $[0.00, 0.50]$ & 2.26 & 0.59 & 17.78 & 1.68 & 1.66 & 11\rule{0pt}{10pt}\\
                                         & $[0.50, 0.73]$ & 0.44 & 7.32 & 12.84 & 4.64 & 4.67 & 12 \\
                                         & $[0.73, 0.87]$ & 1.94 & 9.99 & 321.72 & 0.96 & 0.95 & 15 \\
                                         & $[0.87, 1.20]$ & 3.52 & 7.69 & 927.48 & 2.42 & 2.37 & 17 \\
        \multirow[t]{4}{*}{$[-19, -20]$} & $[0.00, 0.55]$ & 14.03 & 4.07 & 79.00 & 3.49 & 3.46 & 15 \\
                                         & $[0.55, 0.85]$ & 10.28 & 12.86 & 96.65 & 7.82 & 7.10 & 15 \\
                                         & $[0.85, 0.92]$ & 2.97 & 9.98 & 653.35 & 1.66 & 1.27 & 18 \\
                                         & $[0.92, 1.20]$ & 6.02 & 12.78 & 1494.19 & 2.80 & 1.84 & 20 \\
        \multirow[t]{4}{*}{$[-20, -21]$} & $[0.00, 0.72]$ & 137.95 & 8.56 & 436.78 & 24.00 & 23.32 & 18 \\
                                         & $[0.72, 0.90]$ & 77.03 & 54.35 & 1407.20 & 73.66 & 62.11 & 19 \\
                                         & $[0.90, 0.94]$ & 14.36 & 26.47 & 2543.45 & 26.67 & 21.09 & 20 \\
                                         & $[0.94, 1.20]$ & 43.12 & 380.09 & 17572.38 & 50.83 & 33.98 & 20 \\
        \multirow[t]{4}{*}{$[-21, -22]$} & $[0.00, 0.82]$ & 39.26 & 2.97 & 268.66 & 18.78 & 18.76 & 20 \\
                                         & $[0.82, 0.94]$ & 92.94 & 3.95 & 1937.60 & 52.17 & 50.20 & 20 \\
                                         & $[0.94, 0.97]$ & 13.42 & 1.61 & 2899.11 & 13.77 & 13.30 & 20 \\
                                         & $[0.97, 1.20]$ & 4.71 & 8.84 & 4453.61 & 2.50 & 2.22 & 20 \\
        \hline
    \end{tabularx}
\end{table*}

Of the 16 subsamples, the low-luminosity ones ($-20\leq M_{\rm r}\leq -18$) show good agreement between the SDSS measurements and the predictions of the CCMD model. Significant discrepancies appear for subsamples in the $-21\leq M_{\rm r}\leq -20$ luminosity bin, with the CCMD model underpredicting VPFs for cell radius in the range of $5\hinvMpc \lesssim R \lesssim 15\hinvMpc$ (Fig.~\ref{fig:all_VPF}). For the most luminous subsamples ($-22\leq M_{\rm r}\leq -21$), the CCMD model works well, except for the $0.82\leq g-r \leq 0.94$ subsample.

For the analytic models, we find that the LND + Bias model does not fit the SDSS measurements, except for the two blue subsamples in the lowest luminosity bin ($-19\leq M_{\rm r} \leq -18$).  The performance of the GQED model is similar to that of the CCMD model, which generally works well, except for two subsamples in the $-21\leq M_{\rm r} \leq -20$ bin and one subsample in the $-22\leq M_{\rm r} \leq -21$ bin. The NBD model shows overall good agreement with the SDSS measurements, except for the two subsamples in the $-21 \leq M_{\rm r} \leq -20$ luminosity bin with $0.72\leq g-r \leq 0.90$ and $0.94 \leq g-r \leq 1.20$.

In the CCMD mock catalogues used in the above comparison, each central galaxy is assigned the position and velocity of the potential minimum of the host halo, and satellite galaxies are assigned the positions and velocities of randomly chosen dark matter particles inside the halo. Since velocity bias \citep[e.g.][]{Berlind2003,Guo2015BOSS} leads to shifts in redshift-space galaxy positions, it can affect VPF or CIC statistics. We perform a test to examine the velocity bias effect by incorporating it into the mock galaxy catalogues. A random line-of-sight velocity is added to each central galaxy, drawn from a Gaussian distribution with zero mean and standard deviation of $\alpha_c$ times the one-dimensional velocity dispersion of the dark matter particles inside the halo. For each satellite galaxy, its line-of-sight velocity in the rest frame of the host halo is multiplied by a factor $\alpha_s$. We adopt the velocity bias parameters $\alpha_c$ and $\alpha_s$ constrained from modelling redshift-space clustering of SDSS galaxies in \citeauthor{Guo2015} (\citeyear{Guo2015}; see their table 2). Specifically, the values of central galaxy velocity bias $\alpha_c$ for galaxies in the four luminosity bins (from low to high luminosity) are 0.01, 0.29, 0.25, and 0.27, respectively. Those for satellite velocity bias $\alpha_s$ are 0.95, 0.77, 0.84, and 0.97, respectively.

The red dashed curve in each panel of Fig.~\ref{fig:all_VPF} shows the CCMD model with velocity bias incorporated. For most subsamples, the effect is small. However, for the three redder subsamples in the $-21\leq M_{\rm r}\leq -20$ bin, velocity bias leads to a noticeable increase in the VPF amplitude and an improved match to the SDSS measurement. By performing tests with different values of $\alpha_c$ and $\alpha_s$, we find that the above change mainly comes from the satellite velocity bias. With $\alpha_s=0.84$, the line-of-sight distribution of satellite galaxies in redshift space is less stretched, which in turn increases the probability of finding voids. The $\chi^2$ values (Table~\ref{tab:VPF_chi2}) are reduced, while that for the $0.72\leq g-r \leq 0.90$ subsample still remains large ($\sim$62 for d.o.f.$=19$).

The above comparisons reveal that it is difficult for the models to match the SDSS measurements of the luminous subsamples, especially those with $-21\leq M_{\rm r}\leq -20$. We note that these subsamples have the SGW \citep{Gott2005ApJ...624..463G} within their survey volumes. The SGW is in the redshift range of $0.07<z<0.09$, not included in the low-luminosity subsamples. For the most luminous subsamples ($-22\leq M_{\rm r} \leq -21$, with a maximum redshift of 0.165), the SGW occupies only a small fraction of the total volume. With a maximum redshift of 0.110, the subsamples in the $-21\leq M_{\rm r} \leq -20$ bin may be the ones most affected by the SGW. To investigate the effect of the SGW, we set the minimum redshift to be 0.09 for the subsamples in the two luminous bins to remove the SGW and remeasure the VPFs (with number density matched). These measurements are shown as green points with error bars in  Fig.~\ref{fig:all_VPF}. 

For the subsamples with $-21 \leq M_\mathrm{r} \leq -20$, in general, we find that removing the SGW causes the VPF to slightly increase in amplitude for cell radius of $\sim$5--10$\hinvMpc$, except for the bluest subsample with $0.00 \leq g-r \leq 0.72$, for which the VPF is largely unaffected. Note that any change seen here reflects the change in clustering, as the number densities are matched for each subsample.

For the subsamples with $-22 \leq M_\mathrm{r} \leq -21$, with the SGW removed, the VPFs of the two redder subsamples are not affected, while those of the two bluer subsamples decrease slightly for large cells, with radius $\sim$10--20$\hinvMpc$. 

We calculate the $\chi^2$ values for all models with measurements after removing the SGW (see Table~\ref{tab:VPF_chi2_without_wall}), which are shown as dashed lines in Fig.~\ref{fig:VPF_chi2_values}. Removing the SGW does not lead to better agreement between the measurements from the SDSS data and the CCMD mock (except for the subsample of $-22\leq M_{\rm r}\leq -21$ and $0.82\leq g-r\leq 0.94$). In most cases, the deviation of the SDSS measurements from the CCMD model increases (Fig.~\ref{fig:all_VPF}). The lack of an overall monotonic trend in the $\chi^2$ change could be partially caused by noise in the covariance matrices. Note that the CCMD model is constrained by the 2PCFs, with the SGW included for relevant subsamples. The 2PCFs are affected by the rare structures (e.g. see fig. 9 in \citealt{Zehavi2005ApJ...630....1Z} for the impact of the survey volume on measuring the clustering in the data). The 2PCFs predicted by the best-fitting CCMD model for subsamples affected by the SGW show deviations from the SDSS measurements (see fig. 5 in \citealt{Xu2018MNRAS.481.5470X} with $z_{\rm max}$ for each subsample listed in their table A1). Further investigations on the responses of the VPFs and 2PCFs to SGW-like structures and the subsequent effect on the CCMD model are necessary for a fair comparison.

Among the analytic models, we find that removing the SGW improves the NBD model's fit to the SDSS measurements for the $-21\leq M_{\rm r}\leq -20$ subsamples, except for the reddest one. For the GQED model, only the bluest one shows a significant improvement.

Overall, we find that the CCMD model and the GQED and NBD models provide a good description of the SDSS VPF measurements for low-luminosity subsamples, while they have varying levels of difficulty in matching the VPF measurements for high-luminosity subsamples. This cannot be resolved by simply removing the SGW, while velocity bias helps alleviate the mismatch.

\begin{table}
    \centering
    \caption{
    Same as Table~\ref{tab:VPF_chi2}, but for subsamples in the two luminous bins with the SGW excluded.
    }
    \label{tab:VPF_chi2_without_wall}

    \setlength{\tabcolsep}{3pt}

    \begin{tabular}{@{}ccrrrrc@{}}
        \hline
        $M_\mathrm{r}$ (mag) & $g-r$ (mag) & GQED & NBD & \shortstack{LND \\+ Bias} & CCMD & d.o.f.\rule{0pt}{16pt}\\
        \hline
        \multirow[t]{4}{*}{$[-20, -21]$} & $[0.00, 0.72]$ & 79.77 & 8.79 & 315.16 & 25.38 & 19\rule{0pt}{10pt} \\
                                         & $[0.72, 0.90]$ & 88.65 & 14.63 & 1404.00 & 72.73 & 19 \\
                                         & $[0.90, 0.94]$ & 12.30 & 6.69 & 2459.24 & 40.88 & 20 \\
                                         & $[0.94, 1.20]$ & 22.22 & 92.79 & 13130.47 & 70.86 & 20 \\
        \multirow[t]{4}{*}{$[-21, -22]$} & $[0.00, 0.82]$ & 33.61 & 1.17 & 239.89 & 93.96 & 20 \\
                                         & $[0.82, 0.94]$ & 89.80 & 0.58 & 1919.16 & 37.71 & 20 \\
                                         & $[0.94, 0.97]$ & 14.43 & 0.75 & 2797.33 & 16.48 & 20 \\
                                         & $[0.97, 1.20]$ & 4.29 & 8.92 & 3587.55 & 3.84 & 20 \\
        \hline
    \end{tabular}
\end{table}


\subsection{CIC distribution}
\label{subsection:Counts-in-Cells Distribution}

\begin{figure*}
    \centering
    \includegraphics[width=\textwidth]{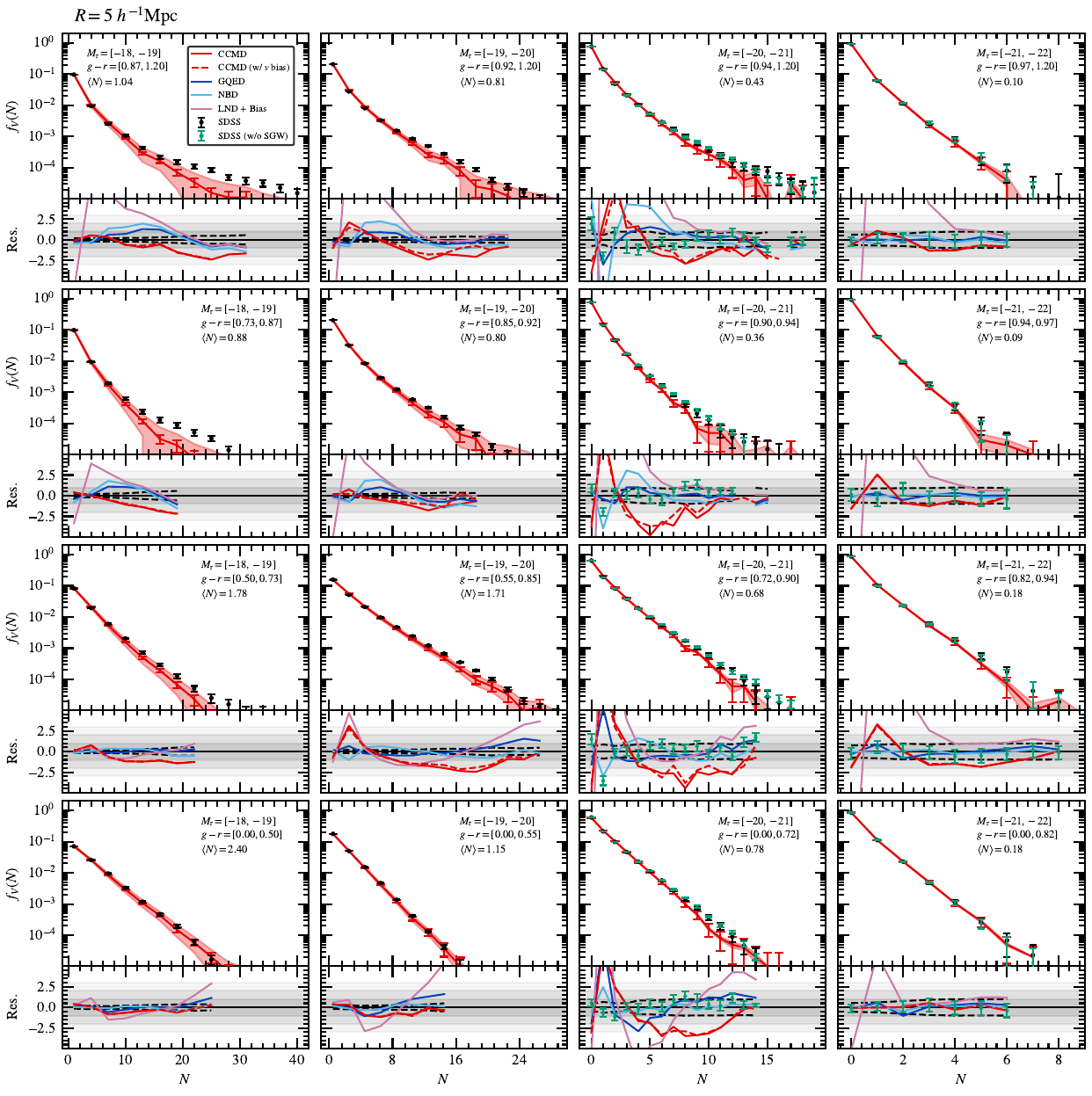}
    \caption{CIC distribution $f_V(N)$ with cell radius $R = 5 \hinvMpc$ for the 16 subsamples. In each panel, we also list the mean number of galaxies in a cell, $\langle N \rangle$. The symbols and curves have the same meanings as in Fig.~\ref{fig:all_VPF}.
    }
    \label{fig:all_f_v(N)_r_5}
\end{figure*}

\begin{figure*}
    \centering
    \includegraphics[width=\textwidth]{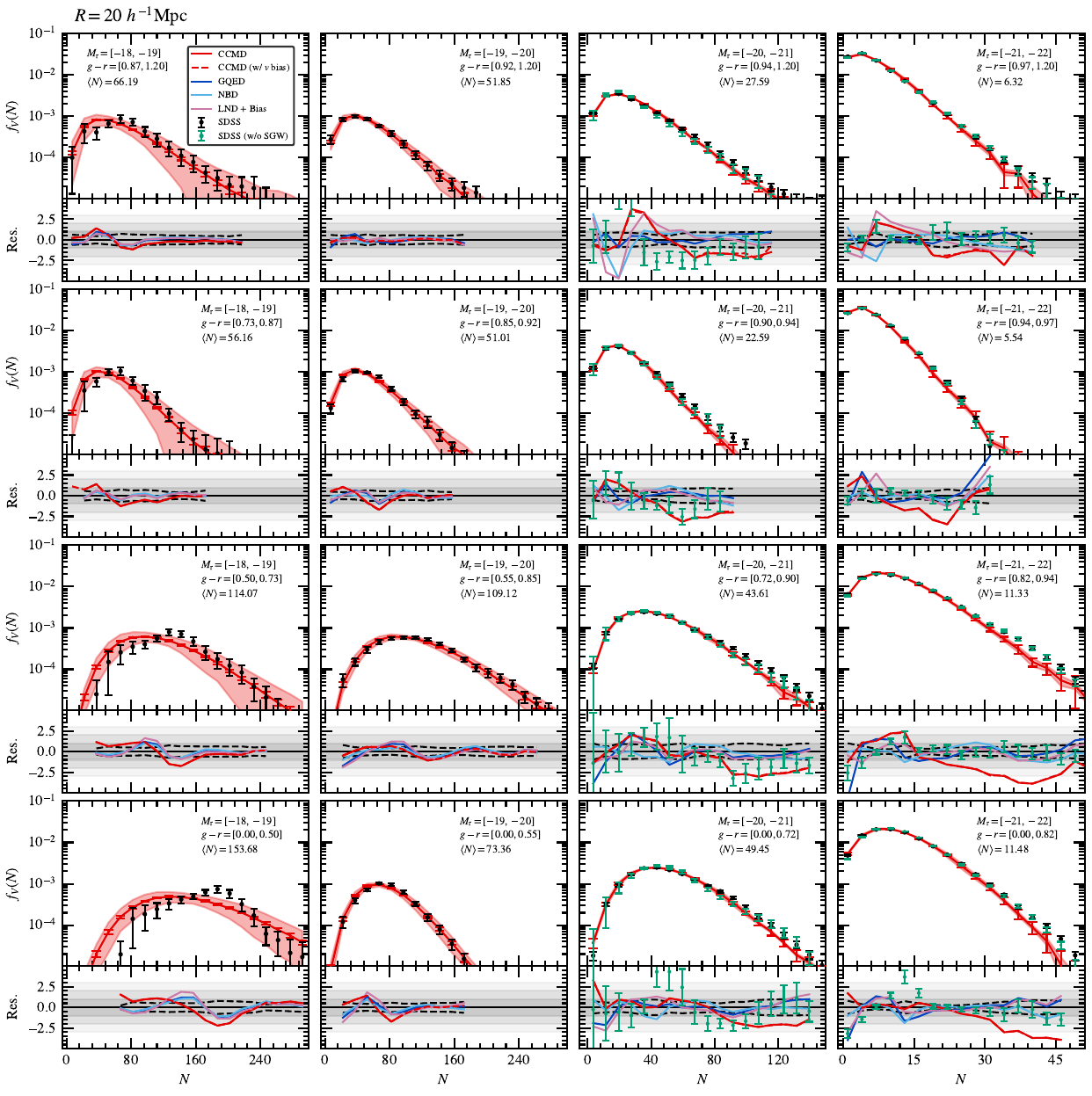}
    \caption{
    Same as Fig.~\ref{fig:all_f_v(N)_r_5} but with cell radius $R=20\hinvMpc$.
    }
    \label{fig:all_f_v(N)_r_20}
\end{figure*}

\begin{figure*}
    \centering
    \begin{subfigure}[b]{0.45\textwidth}
        \includegraphics[width=\linewidth]{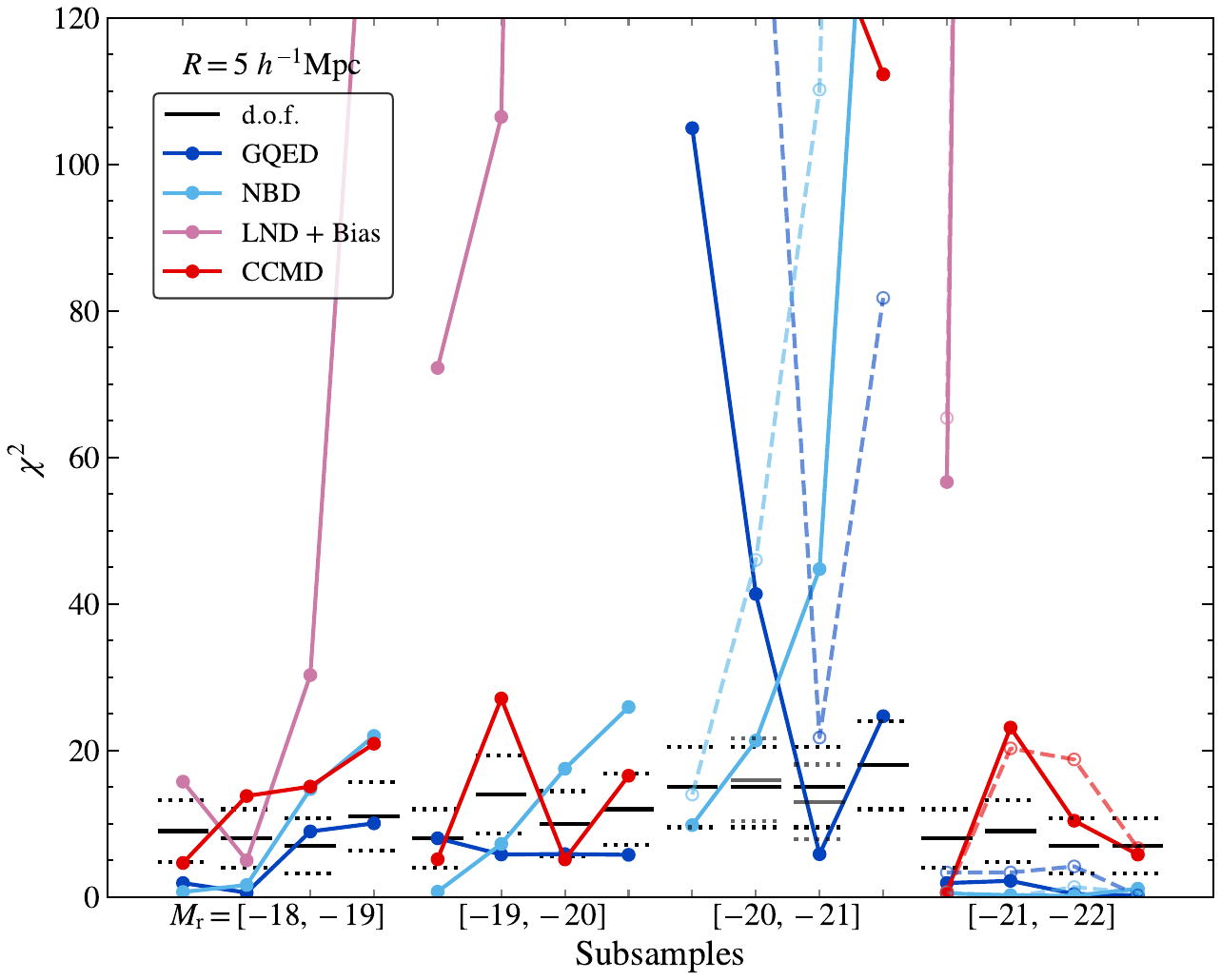}
    \end{subfigure}
    \begin{subfigure}[b]{0.45\textwidth}
        \includegraphics[width=\linewidth]{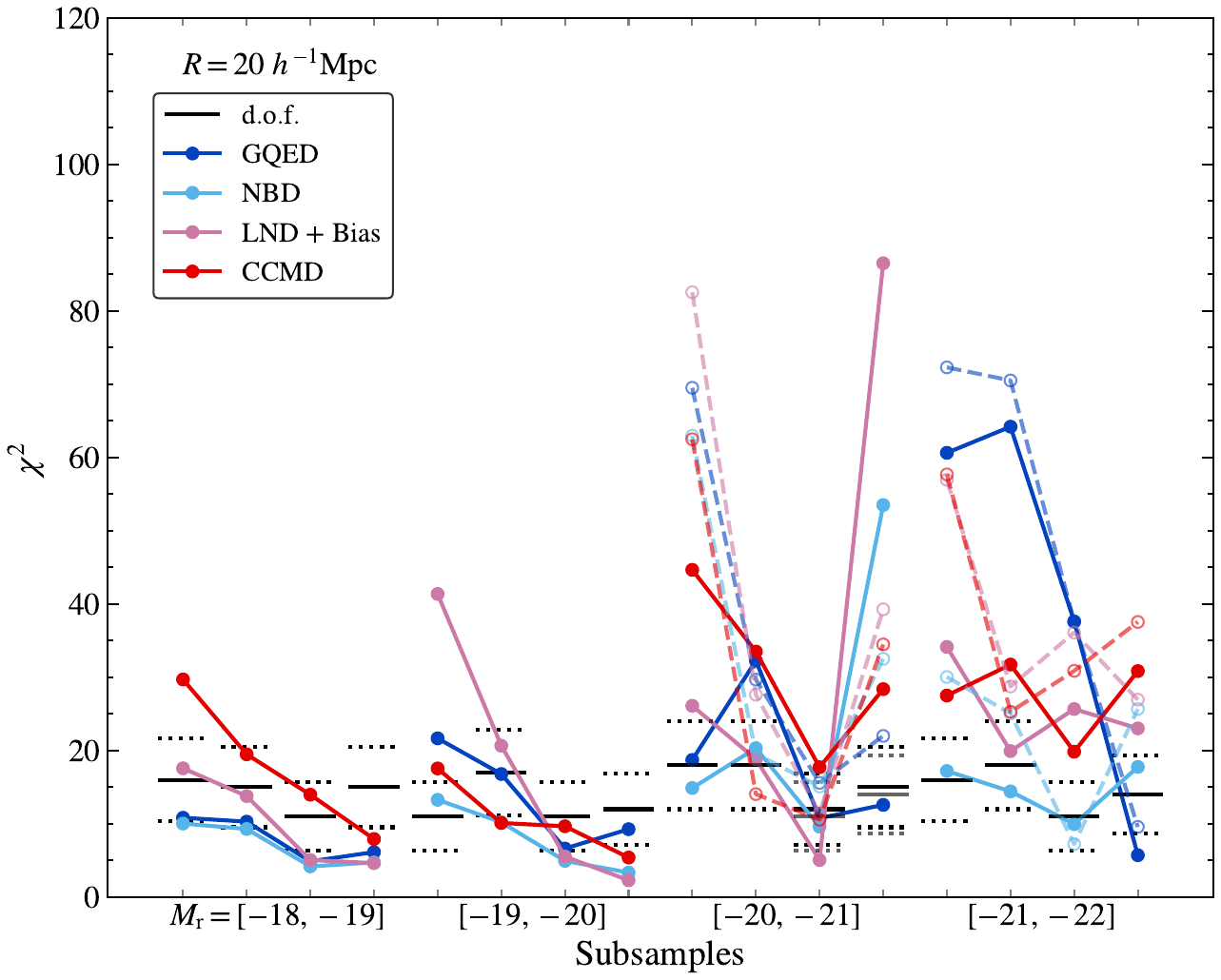}
    \end{subfigure}
    \caption{
    Same as Fig.~\ref{fig:VPF_chi2_values} but for the CIC $f_V(N)$ measurements with cell radius $R=5\hinvMpc$ (left) and $20\hinvMpc$ (right).
    }
    \label{fig:f_v_(N)_chi2_values}
\end{figure*}

We show the distribution of galaxy count, $f_V(N)$, of the 16 subsamples in Figs~\ref{fig:all_f_v(N)_r_5} and \ref{fig:all_f_v(N)_r_20}, for cell radius $R = 5 \hinvMpc$ and $20 \hinvMpc$, respectively. We follow Section~\ref{subsection:Void Probability Distribution} to construct the covariance matrices and to compare the SDSS measurements with the CCMD and analytic models. The $\chi^2$ values are calculated for different models for measurements with and without the SGW, which are listed in Tables~\ref{tab:f_v(N)_chi2} and \ref{tab:f_v(N)_chi2_without_SGW} and shown in Fig.~\ref{fig:f_v_(N)_chi2_values}.

With $R=5\hinvMpc$, the mean number of galaxies per cell for each of the 16 subsamples typically has $\bar{N}\lesssim 2$. With the binning scheme in Fig.~\ref{fig:all_f_v(N)_r_5}, $f_V(N)$ is similar to an exponential function. The CCMD model provides a good description of the SDSS measurements, except for the subsamples in the $-21\leq M_{\rm r}\leq -20$ luminosity bin (see the left panel of Fig.~\ref{fig:f_v_(N)_chi2_values}). For subsamples in this luminosity bin, the disagreements are mainly at $N=0$ (i.e. VPF) and at $N=1$, with the CCMD model underestimating the VPF and overestimating $f_V(N=1)$. Incorporating velocity bias alleviates the disagreements between the CCMD model and the SDSS measurements (Fig.~\ref{fig:all_f_v(N)_r_5}), with a modest reduction in $\chi^2$ values (Table~\ref{tab:f_v(N)_chi2_without_SGW}).

Among the three analytic models, the LND + Bias model fails to match the SDSS measurements with $R=5\hinvMpc$, except for the two blue subsamples in the lowest luminosity bin. The GQED and NBD models agree with the measurements, but not for the subsamples in the $-21\leq M_{\rm r}\leq -20$ luminosity bin. Removing the SGW causes $f_V(N)$ to slightly shift toward lower $N$ for the three redder subsamples in the $-21\leq M_{\rm r}\leq -20$ luminosity bin, with little change for the subsamples in the $-22\leq M_{\rm r}\leq -21$ luminosity bin. As with the VPF case, excluding the SGW does not improve the agreement between the CCMD, GQED, and NBD models and the SDSS $f_V(N)$ measurements. 

With $R=20\hinvMpc$, SDSS $f_V(N)$ measurements show Poisson-like distributions with mean ranging from a few (for subsamples in the most luminous bin) to tens or above one hundred (Fig.~\ref{fig:all_f_v(N)_r_20}). The effect of removing the SGW on $f_V(N)$ is similar to the case with $R=5\hinvMpc$.

The CCMD model and all three analytic models provide reasonable descriptions of SDSS $f_V(N)$ measurements with $R=20\hinvMpc$ for most subsamples (see the $\chi^2$ values in the right panel of Fig.~\ref{fig:f_v_(N)_chi2_values}). Subsamples with the lowest luminosity ($-19\leq M_{\rm r}\leq -18$) have the smallest volume ($\sim 1.6\times 10^6 h^{-3}{\rm Mpc}^3$), which is less than 50 times the cell volume. Therefore, their CIC measurements show the largest sample variance effect, with shapes deviating from the CCMD model predictions but still consistent within the uncertainties (see the leftmost column of Fig.~\ref{fig:all_f_v(N)_r_20}, especially the two blue subsamples). For the CCMD model, the least satisfied cases are the bluest and reddest subsamples in the $-21\leq M_{\rm r}\leq -20$ luminosity bin, with $\chi^2 \simeq$ 45 and 133 for d.o.f. = 18 and 16, respectively. For most subsamples in the two luminous bins, removing the SGW does not improve the agreement between models and the SDSS measurements, and in many cases the level of agreement is reduced (dashed lines in the right panel of Fig.~\ref{fig:f_v_(N)_chi2_values}). With the large cell size, velocity bias has virtually no effect on the CCMD model predictions (Fig.~\ref{fig:all_f_v(N)_r_20} and Table~\ref{tab:f_v(N)_chi2_without_SGW}).

\begin{table*}
    \centering
    \caption{
    Values of $\chi^2$ of different models in describing the CIC $f_V(N)$ measurements of 16 subsamples with cell radius $R=5\hinvMpc$ and $20\hinvMpc$. For the two luminous bins, the subsamples include the SGW. The d.o.f. for each subsample and each cell radius are also listed. 
    }
    \label{tab:f_v(N)_chi2}
    \setlength{\tabcolsep}{5pt}
    \begin{tabularx}{17.75cm}{ccrrrrrcrrrrrc}
        \hline
        & & \multicolumn{6}{c}{$R = 5 \hinvMpc$} & \multicolumn{6}{c}{$R = 20 \hinvMpc$} \rule{0pt}{10pt} \\
        \hline
        $M_\mathrm{r}$ (mag) & $g-r$ (mag) & GQED & NBD & \shortstack{LND\\+ Bias} & CCMD & \shortstack{CCMD\\(w/ $v$ bias)} & d.o.f. & GQED & NBD & \shortstack{LND\\+ Bias} & CCMD & \shortstack{CCMD\\(w/ $v$ bias)} & d.o.f.\rule{0pt}{17pt}\\
        \hline
        \multirow[t]{4}{*}{$[-18, -19]$} & $[0.00, 0.50]$ & 1.92 & 0.70 & 15.76 & 4.66 & 4.54 & 9 & 10.83 & 10.04 & 17.58 & 29.71 & 29.76 & 16\rule{0pt}{10pt}\\
                                         & $[0.50, 0.73]$ & 0.61 & 1.65 & 5.01 & 13.80 & 13.44 & 8 & 10.30 & 9.30 & 13.79 & 19.49 & 19.51 & 15\\
                                         & $[0.73, 0.87]$ & 8.96 & 14.75 & 30.30 & 15.09 & 14.63 & 7 & 4.88 & 4.19 & 5.06 & 13.98 & 14.12 & 11\\
                                         & $[0.87, 1.20]$ & 10.07 & 22.00 & 159.16 & 20.94 & 21.27 & 11 & 6.14 & 4.74 & 4.64 & 7.93 & 8.36 & 15\\
        \multirow[t]{4}{*}{$[-19, -20]$} & $[0.00, 0.55]$ & 8.03 & 0.74 & 72.23 & 5.17 & 5.88 & 8 & 21.67 & 13.28 & 41.38 & 17.56 & 17.53 & 11\\
                                         & $[0.55, 0.85]$ & 5.82 & 7.26 & 106.50 & 27.11 & 20.73 & 14 & 16.80 & 10.22 & 20.67 & 10.12 & 9.97 & 17\\
                                         & $[0.85, 0.92]$ & 5.87 & 17.52 & 529.31 & 5.15 & 6.42 & 10 & 6.63 & 4.94 & 5.54 & 9.67 & 10.62 & 11\\
                                         & $[0.92, 1.20]$ & 5.77 & 25.94 & 1452.26 & 16.57 & 8.20 & 12 & 9.27 & 3.35 & 2.30 & 5.40 & 4.83 & 12\\
        \multirow[t]{4}{*}{$[-20, -21]$} & $[0.00, 0.72]$ & 104.96 & 9.83 & 613.27 & 161.74 & 151.66 & 15 & 18.72 & 14.89 & 26.11 & 44.67 & 44.48 & 18\\
                                         & $[0.72, 0.90]$ & 41.37 & 21.36 & 1199.98 & 198.92 & 164.24 & 15 & 32.27 & 20.33 & 18.83 & 33.51 & 31.99 & 18\\
                                         & $[0.90, 0.94]$ & 5.86 & 44.77 & 1846.86 & 134.29 & 108.76 & 15 & 10.77 & 9.62 & 5.06 & 17.73 & 17.11 & 12\\
                                         & $[0.94, 1.20]$ & 24.69 & 179.96 & 4994.73 & 112.31 & 69.32 & 18 & 12.59 & 53.52 & 86.52 & 28.39 & 29.23 & 15\\
        \multirow[t]{4}{*}{$[-21, -22]$} & $[0.00, 0.82]$ & 1.92 & 0.49 & 56.66 & 0.73 & 0.73 & 8 & 60.64 & 17.21 & 34.15 & 27.51 & 27.19 & 16\\
                                         & $[0.82, 0.94]$ & 2.22 & 0.26 & 705.39 & 23.14 & 21.41 & 9 & 64.21 & 14.43 & 19.93 & 31.74 & 31.30 & 18\\
                                         & $[0.94, 0.97]$ & 0.44 & 0.16 & 929.56 & 10.43 & 9.73 & 7 & 37.62 & 9.96 & 25.67 & 19.85 & 20.38 & 11\\
                                         & $[0.97, 1.20]$ & 0.18 & 1.15 & 1359.66 & 5.81 & 5.09 & 7 & 5.72 & 17.77 & 23.03 & 30.85 & 30.38 & 14\\
        \hline
    \end{tabularx}
\end{table*}

\begin{table*}
    \centering
    \caption{
    Same as Table~\ref{tab:f_v(N)_chi2}, but for subsamples in the two luminous bins with the SGW excluded.
    }
    \label{tab:f_v(N)_chi2_without_SGW}
    \begin{tabularx}{16.5cm}{ccrrrrcrrrrc}
        \hline
        & & \multicolumn{5}{c}{$R = 5 \hinvMpc$} & \multicolumn{5}{c}{$R = 20 \hinvMpc$} \rule{0pt}{10pt} \\
        \hline
        $M_\mathrm{r}$ (mag) & $g-r$ (mag) & GQED & NBD & LND + Bias & CCMD & d.o.f. & GQED & NBD & LND + Bias & CCMD & d.o.f.\rule{0pt}{10pt}\\
        \hline
        \multirow[t]{4}{*}{$[-20, -21]$} & $[0.00, 0.72]$ & 165.72 & 14.00 & 1034.15 & 317.87 & 15 & 69.52 & 62.97 & 82.56 & 62.50 & 18\rule{0pt}{10pt}\\
                                         & $[0.72, 0.90]$ & 159.49 & 46.02 & 2948.65 & 617.74 & 16 & 29.69 & 19.56 & 27.63 & 14.06 & 18\\
                                         & $[0.90, 0.94]$ & 21.78 & 110.22 & 1750.12 & 358.27 & 13 & 15.59 & 15.09 & 11.43 & 10.57 & 11\\
                                         & $[0.94, 1.20]$ & 81.78 & 440.70 & 7823.12 & 227.17 & 18 & 22.00 & 32.52 & 39.28 & 34.49 & 14\\
        \multirow[t]{4}{*}{$[-21, -22]$} & $[0.00, 0.82]$ & 3.35 & 0.66 & 65.39 & 0.44 & 8 & 72.31 & 30.04 & 56.95 & 57.70 & 16\\
                                         & $[0.82, 0.94]$ & 3.36 & 0.03 & 708.18 & 20.30 & 9 & 70.52 & 25.07 & 28.75 & 25.30 & 18\\
                                         & $[0.94, 0.97]$ & 4.19 & 1.37 & 1143.28 & 18.79 & 7 & 37.58 & 7.23 & 36.11 & 30.88 & 11\\
                                         & $[0.97, 1.20]$ & 0.23 & 0.58 & 1803.25 & 6.68 & 7 & 9.58 & 25.67 & 26.95 & 37.56 & 14\\
        \hline
    \end{tabularx}
\end{table*}

\subsection{CIC distribution, halo mass function, and halo occupation function}
\label{subsection:Counts-in-Cells and Halo Occupation}

\begin{figure*}
    \centering
    \includegraphics[width=\textwidth]{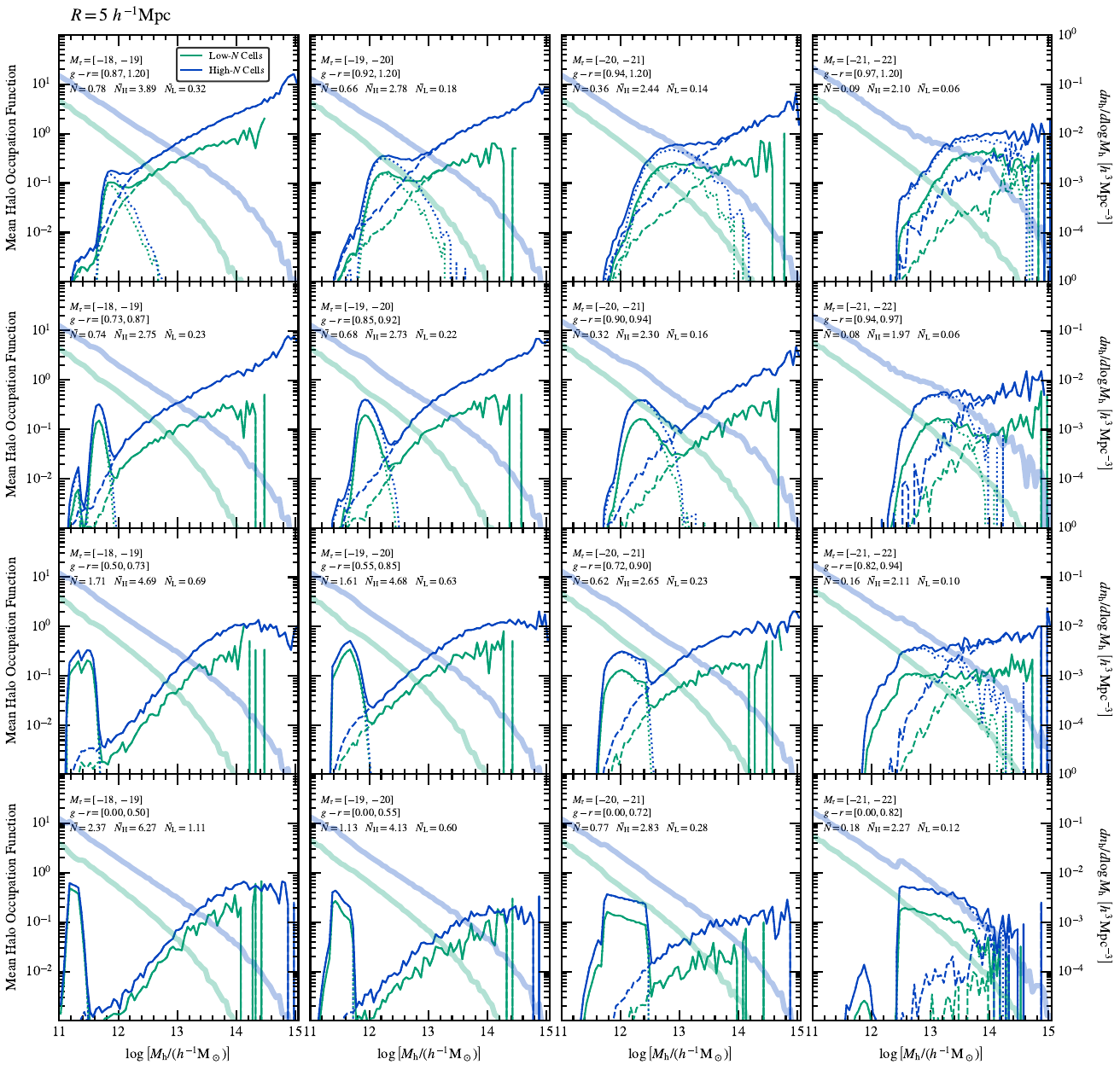}
    \caption{
    Halo mass function and mean halo occupation function in cells of radius $R=5\hinvMpc$ for the 16 subsamples. In each panel, the transparent curves represent the halo mass functions (as labelled in the rightmost axes), while the opaque curves represent the mean halo occupation functions (as labelled in the leftmost axes). The green (blue) curves come from cells that contain a small (large) number of galaxies. The mean numbers of galaxies in all cells, high-$N$ cells, and low-$N$ cells are listed as $\bar{N}$, $\bar{N}_{\rm H}$, and $\bar{N}_{\rm L}$ (see the text for details). For mean halo occupation functions, the dotted and dashed curves are contributions from central and satellite galaxies, respectively.
    }
    \label{fig:halo_occupation_r5}
\end{figure*}

\begin{figure*}
    \centering
    \includegraphics[width=\textwidth]{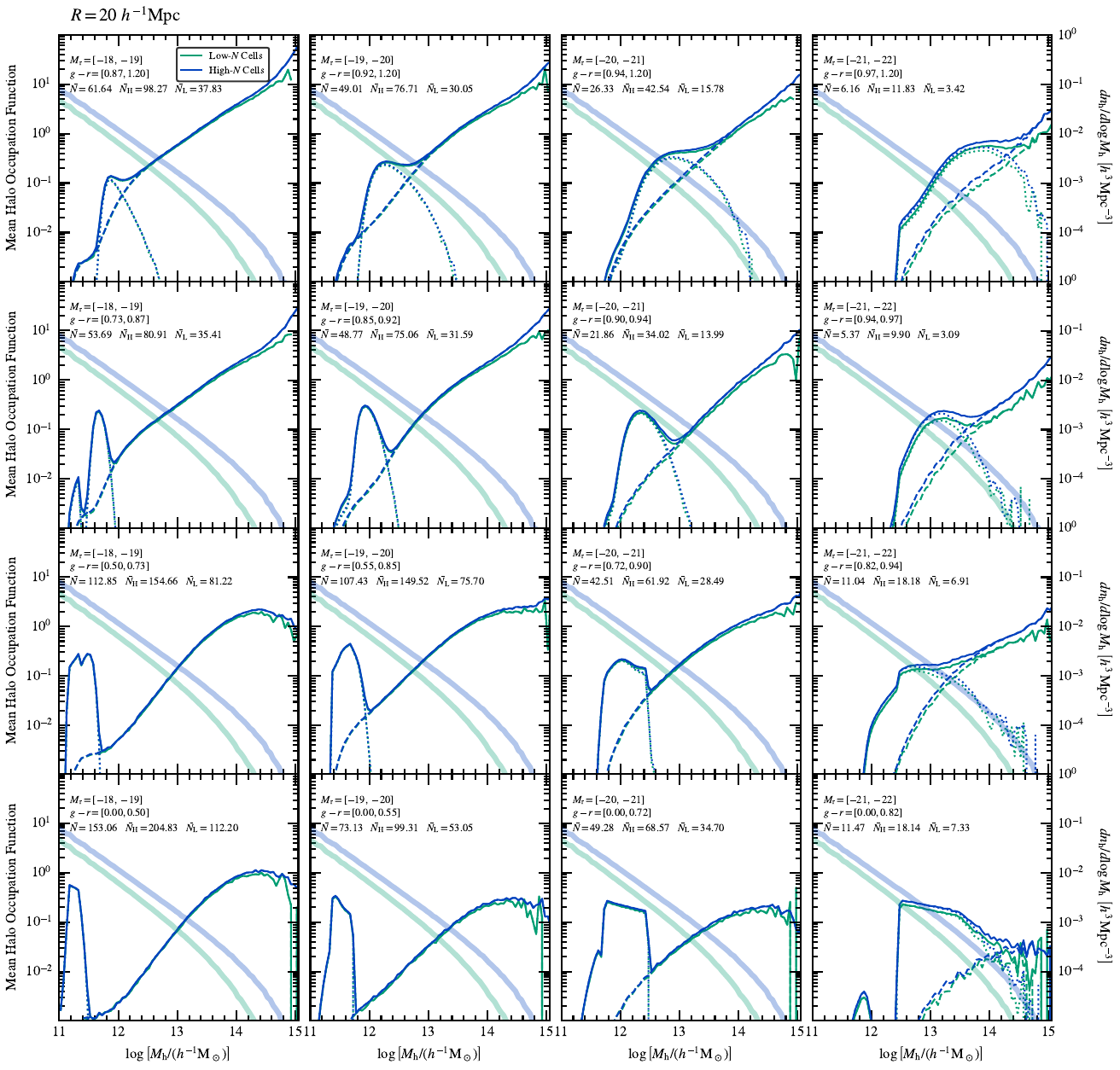}
    \caption{
    Same as Fig.~\ref{fig:halo_occupation_r5} but with cell radius $R=20\hinvMpc$.
    }
    \label{fig:halo_occupation_r20}
\end{figure*}

For CIC statistics, at a fixed cell size, the number of galaxies contained in a cell varies, depending on the position of the cell. Within the halo model, the CIC variation can have two contributions. First, the halo abundance (i.e. halo mass function) in a cell can vary, reflecting the underlying matter density fluctuation or the environment fluctuation. Second, the number of galaxies per halo (i.e. halo occupation function) in a cell can vary. In the CCMD model, the statistical distribution of galaxies, in terms of luminosity and colour, is assumed to depend only on the halo mass and not on the halo environment. For example, for a given subsample, at fixed halo mass, the number of satellite galaxies follows a Poisson distribution with the mean depending only on halo mass. It is possible that the fluctuations in the halo occupation distribution can be captured by the CIC statistics.

Using the CCMD mocks, we can explore the contributions from the variations in the halo mass function and the halo occupation function to help our understanding of the CIC statistics. For each cell radius, we divide the cells into two groups for each subsample by comparing the number of galaxies in the cells to the mean number $\bar{N}$ of galaxies per cell. The first group consists of cells with a low number of galaxies (low-$N$), with $N \leq \lceil\bar{N}\rceil$, where $\lceil \rceil$ represents the ceiling function. The second group consists of cells with a high number of galaxies (high-$N$), with $N > \lceil\bar{N}\rceil$. For each cell, we use the halo catalogue (in redshift space) to identify haloes whose centres lie within the cell. In order to calculate the halo occupation function (mean number of galaxies per halo as a function of halo mass), we identify galaxies within the cell using the galaxy mock catalogue (in redshift space) of each subsample. 

There is a small fraction of galaxies inside the cell, but with centres of their host haloes outside the cell. These galaxies are not included in our calculation of the halo occupation function, and given the small fraction, excluding them would not affect our main results. For example, for cell radius $R = 5 \hinvMpc$, these galaxies represent $\sim 3$ per cent of all galaxies in the blue subsamples and $\sim 18$ per cent in the red subsamples, and for cell radius $R = 20 \hinvMpc$, the fractions become $\sim 1$ per cent and $\sim 5$ per cent in the blue and red subsamples, respectively. The difference in the percentages between the blue and red subsamples results from higher satellite fractions in red subsamples, and the small-scale redshift-space distortions have a larger effect on the CIC statistics. 

With haloes and galaxies inside cells identified for each subsample, for either group of cells (low-$N$ or high-$N$), the corresponding halo mass function is computed as the mean halo number density as a function of halo mass. For each halo mass bin, we collect galaxies residing in such haloes and compute the mean halo occupation function as the mean number of galaxies per halo as a function of halo mass.

We show the halo mass functions and the halo occupation functions of the 16 subsamples in Fig.~\ref{fig:halo_occupation_r5} and Fig.~\ref{fig:halo_occupation_r20}, which correspond to cell radius $R = 5 \hinvMpc$ and $20 \hinvMpc$, respectively. The green curves correspond to low-$N$ cells, while the blue curves correspond to high-$N$ cells. The transparent thick curves represent the halo mass function, while the opaque thin curves represent the mean halo occupation function, separated into contributions from central galaxies (dotted curves) and satellite galaxies (dashed curves). In each panel, we also list $\bar{N}$, $\bar{N}_{\rm H}$, and $\bar{N}_{\rm L}$, the mean number of galaxies in all cells, high-$N$ cells, and low-$N$ cells, respectively. The mean number of galaxies in a cell is the product of the mean halo mass function and the mean halo occupation function, integrated over halo mass. 
As an approximate sanity check, the ratio of $\bar{N}_{\rm H}$ to $\bar{N}_{\rm L}$ can be compared to the product of the ratio of the mean halo mass functions of high-$N$ and low-$N$ cells and that of the mean halo occupation functions, and we find general agreement.

With cells of $R = 5 \hinvMpc$ (Fig.~\ref{fig:halo_occupation_r5}), we find that the halo mass function in high-$N$ cells is higher than in low-$N$ cells by a factor of $\sim 6.3$ on average, indicating that high-$N$ cells reside in regions overdense in matter. With cells of $R = 20 \hinvMpc$ (Fig.~\ref{fig:halo_occupation_r20}), the matter density fluctuation becomes smaller. Therefore, the separation between the halo mass functions in high-$N$ and low-$N$ cells becomes smaller, with an average ratio of $\sim 3.6$.

In Fig.~\ref{fig:halo_occupation_r5}, with $R=5\hinvMpc$, we find that the mean halo occupation function in high-$N$ cells is higher than in low-$N$ cells by a factor of $\sim 3.4$ on average. The results seem to contradict the assumption in the CCMD model that the halo occupation distribution depends only on halo mass, not environment. The key to understanding the result is that the halo occupation has fluctuations around the mean occupation function:  for haloes of a given mass, the occupation number of central galaxies can be either zero or one, and that of satellite galaxies follows a Poisson distribution. With $R=5\hinvMpc$, the mean number of galaxies per cell is on the order of unity. Therefore, the counts are dominated by galaxies in one or a few haloes, and fluctuations in halo occupation can have a large contribution to the variation in the counts in cells. That is, high-$N$ (low-$N$) cells tend to pick up haloes whose occupation numbers happen to fluctuate high (low), leading to an apparently high (low) occupation function. 

With $R=20\hinvMpc$ (Fig.~\ref{fig:halo_occupation_r20}), we see that the mean occupation functions in the high-$N$ and low-$N$ cells are nearly indistinguishable, except for the high-mass end and those of the most luminous subsamples. For the most luminous subsamples, the mean count per cell is about a few, and the occupation fluctuation can still have a noticeable contribution to the cell count variation. For lower-luminosity subsamples, the mean galaxy count (or the mean halo count) per cell is about tens to hundreds, which averages out occupation fluctuations and leads to similar occupation functions in low- and high-$N$ cells. Given the exponential drop in the abundance of the most massive haloes, there is a large fluctuation in the number of high-mass haloes per cell, which explains why the occupation fluctuation at the extremely high-mass end can still show up.

Combining the above results, we see that for high-$N$ cells with radius $R = 5 \hinvMpc$, both the halo mass function and the halo occupation function fluctuate high. For cells with radius $R = 20 \hinvMpc$, the fluctuation in the halo mass function dominates. In general, if the mean galaxy count per cell is low (e.g. order of unity or lower), the variation in counts in cells is contributed by fluctuations in both the halo mass function and the halo occupation distribution. If the mean galaxy count per cell is high (e.g. more than a few tens), the variation in counts in cells is dominated by the fluctuation in the halo mass function.

The halo occupation functions shown in Figs~\ref{fig:halo_occupation_r5} and \ref{fig:halo_occupation_r20} also manifest several aspects of the CCMD model. As luminosity increases, the mean halo occupation function of galaxies shifts toward the high halo mass end, reflecting that more luminous galaxies tend to reside in more massive haloes. At a fixed luminosity bin (i.e. for a given column in the plots), from the blue to red subsamples, the mean halo occupation function of the central galaxies, represented by dotted curves, shifts toward higher halo mass. This is a reflection that in the CCMD model redder central galaxies on average reside in more massive haloes. In addition, the relative contribution of the satellite to central galaxies increases from the blue to red subsamples, resulting from the higher satellite fraction in redder galaxies. In a few subsamples, we see two clear peaks in the mean occupation function of central galaxies, manifesting the contributions from the two pseudo-colour central populations (pseudo-blue and pseudo-red) assumed in the CCMD model.

As a whole, from analysing the CIC statistics using the CCMD mocks, we find that the galaxy number count in a cell is not necessarily solely determined by the matter environment in the cell. For cells of low mean count (order of unity or below), the cell count variation comes from fluctuations both in the matter environment, which causes fluctuations in the halo mass function, and in the occupation distribution of galaxies. For cells of high mean count, the cell count variation is largely due to fluctuations in the mass environment (halo mass function). 

The result has important implications for theoretical studies of galaxy assembly bias from galaxy formation models. In such  studies, it is useful to investigate the environment dependence of the mean halo occupation function of galaxies \citep[e.g.][]{Berlind2003,Hadzhiyska2021}, since a lack of environment dependence suggests no galaxy assembly bias. Based on our CIC analysis with the CCMD mocks, the environment (e.g. overdensity and tidal field) should be defined through the underlying matter distribution or halo distribution. If the environment is defined through the galaxy distribution (e.g. galaxy density from counts in small cells), an apparent variation in the occupation function would be revealed, which, if attributed to the environment-dependent halo occupation, can cause an incorrect inference about galaxy assembly bias. In observational studies, it is common to define the environment based on the galaxy density field, since the matter or halo density field is not directly observable. Depending on the statistics to be investigated, one should disentangle the effect from fluctuations of the halo occupation and that from galaxy assembly bias.


\section{Summary and Discussion}
\label{section:Summary and Discussion}

In this paper, we study CIC statistics as a function of galaxy colour and luminosity. We divide the SDSS main sample into 16 colour--magnitude bins. For each colour--magnitude subsample, we perform CIC measurements with spherical cells of different radii. We compare the measurements with models, including the halo-based CCMD model and three analytic models. With the CCMD model, we also study the connection between the CIC statistics and the variations in halo abundance and halo occupation distribution.

For CIC measurements with SDSS data, we apply a method of using random catalogues to distribute cells entirely within the volume of the data sample, accounting for the edge of the footprint and the masked regions within the footprint. The measured VPFs show the expected trend that more luminous and redder galaxies are more strongly clustered. The measured CIC distribution function $f_V(N)$ of each subsample follows a Poisson-like distribution. 

We compare the SDSS VPF and CIC measurements with the CCMD model and three analytic models, GQED, NBD, and LND + Bias. The CCMD model provides a reasonable match to the SDSS measurements, except for the VPF and $R=5\hinvMpc$ CIC measurements for the $-21\leq M_{\rm r}\leq -20$ subsamples. In general, both the GQED and NBD models provide a good description of the measurements except for a few subsamples in the two luminous bins. The LND + Bias model does not fit the SDSS measurements well on small scales (i.e. with a small cell radius $R$). Our results on the effectiveness of the three analytic models in describing CIC statistics largely agree with previous work (see \citealt{Hurtado-Gil2017A&A...601A..40H} and discussions therein), while we extend the study to the colour- and luminosity-dependent CIC statistics. 

The CIC measurements can be affected by rare structures in the survey volume, and we investigate the effect of the SGW, which are in the survey volumes of subsamples in the two luminous bins. We find that, in general, excluding the SGW does not improve the match between models and measurements. Therefore, further investigations are necessary to study the mismatch between the models and CIC measurements (especially for cells of small sizes). We find that incorporating velocity bias can alleviate the tension between the CCMD model and the SDSS VPF and $R=5\hinvMpc$ CIC measurements in the $-21\leq M_{\rm r}\leq -20$ bin. In this luminosity bin, the velocity bias constraints we adopt are from HOD modelling of the redshift-space 2PCFs \citep{Guo2015}, which are affected by the SGW. As the model itself does not account for the SGW effect on halo distribution (e.g. halo mass function and clustering), HOD and velocity bias constraints can be driven to compensate for such inaccurate model assumptions on halo distribution. Therefore,  the effect of rare structures on CIC statistics or generally on the clustering statistics (e.g. 2PCF; \citealt{Zehavi2005ApJ...630....1Z}) needs to be carefully taken into account in modelling (e.g. with the CCMD model and velocity bias constraints) or in comparison between model and observation. The effect also needs to be addressed for robust constraints on galaxy assembly bias.

With the help of the CCMD mocks, we investigate the cause of variations in galaxy counts in cells. For small cells or cells with low mean galaxy count (order of unity or below), the variation reflects the effect both from the fluctuation in the matter environment and that in the halo occupation distribution. For large cells or cells with high mean galaxy count, the environment effect dominates. In theoretical studies of galaxy assembly bias, environment should be defined using the matter or halo density field. In observational studies of galaxy assembly bias, if the environment is defined through small-scale galaxy distributions, we need to separate the effect of statistical fluctuations in the halo occupation distribution from that of galaxy assembly bias.

In our work, the CCMD mocks are constructed using parameters inferred from fitting the colour- and luminosity-dependent 2PCFs. For better comparison with the CIC statistics, it would be useful to jointly fit the CIC measurements to constrain the parameters \citep[e.g.][]{Wang2019}. Furthermore, galaxy clustering measurements, including CIC statistics, can be improved using data from larger galaxy surveys, such as the Dark Energy Spectroscopic Instrument (DESI; \citealt{DESICollaboration2026AJ....171..285D}) survey, which can tighten model parameter constraints and lead to stringent model tests.


\section*{Acknowledgements}

We thank Xiang Hu for his help with the jackknife samples of the SDSS galaxies and Haojie Xu for helpful comments and discussions. We thank the referee for the constructive comments. The support and resources from the Center for High Performance Computing at the University of Utah are gratefully acknowledged.


\section*{Data Availability}

NYU-VAGC DR7 {\tt dr72bright1} data set and random catalogues are publicly available at \url{ https://sdss.physics.nyu.edu/vagc/}. CCMD galaxy and halo mock catalogues are publicly available at \url{https://zhengzheng-astro.github.io/data.html}. The data underlying this article will be shared upon reasonable request
to the authors.

\balance
\bibliographystyle{mnras}
\bibliography{references}


\bsp	
\label{lastpage}
\end{document}